# Cyclical Variational Bayes Monte Carlo for Efficient Multi-Modal Posterior Distributions Evaluation

Felipe Igea[a*], Alice Cicirello[a,b]

[a] Department of Engineering Science

University of Oxford

Parks Road, Oxford OX1 3PJ, UK

[b] Faculty of Civil Engineering and Geosciences, Department of Engineering Structures,

Section of Mechanics and Physics of Structures (MPS)

Delft University of Technology, Stevinweg 1, Delft 2628, NL

**Abstract**

Multi-modal distributions of some physics-based model parameters are often encountered in engineering due to different situations such as a change in some environmental conditions, and the presence of some types of damage and non-linearity. In statistical model updating, for locally identifiable parameters, it can be anticipated that multi-modal posterior distributions would be found. The full characterization of these multi-modal distributions is important as methodologies for structural condition monitoring in structures are frequently based in the comparison of the damaged and healthy models of the structure. The characterization of posterior multi-modal distributions using state-of-the-art sampling techniques would require a large number of simulations of expensive-to-run physics-based models. Therefore, when a limited number of simulations can be run, as it often occurs in engineering, the traditional sampling techniques would not be able to capture accurately the multi-modal distributions. This could potentially lead to large numerical errors when assessing the performance of an engineering structure under uncertainty.

Therefore, an approach is proposed for drastically reducing the number of models runs while yielding accurate estimates of highly multi-modal posterior distributions. This

[*]Corresponding author: Mr. Felipe Igea

*E-mail address: felipe.igea@eng.ox.ac.uk*

approach introduces a cyclical annealing schedule into the Variational Bayes Monte Carlo (VBMC) method to improve the algorithm's phase of exploration and the finding of high probability areas in the multi-modal posteriors throughout the different cycles.

Three numerical and one experimental investigations are used to compare the proposed cyclical VBMC with the standard VBMC algorithm, the monotonic VBMC and the Transitional Ensemble Markov Chain Monte Carlo (TEMCMC). It is shown that the standard VBMC fails in capturing multi-modal posteriors as it is unable to escape already found regions of high posterior density. In the presence of highly multi-modal posteriors, the proposed cyclical VBMC algorithm outperforms all the other approaches in terms of accuracy of the resulting posterior, and number of model runs required.

## 1 Introduction

Statistical model updating techniques are frequently used in engineering to quantify the inherent variability of some uncertain latent parameters, or to identify the unknown values of latent parameters used in physics-based models in the light of measurements of some observable quantities [1]. These statistically updated models can then be used to evaluate the behaviour of an engineering system under uncertainties. For example, the statistically updated model can be used for assessing the performance of a structure with uncertain input parameters under various loading and environmental conditions, and/or to assess the remaining useful life of such structure [1–4].

Multi-modality on the distributions of some physics-based model's parameters of an engineering system is frequently found [5]. Multi-modality in the latent parameters of a physics-based model may be encountered due to different reasons such as changes in the environmental conditions [6] (e.g., change in stiffness due to varying temperature), the presence of some types of damage and non-linearity [7] (e.g., change in localised stiffness due to opening and closing of a crack). In statistical model updating it is also expected to find multi-modal distributions for locally identifiable problems [8] (e.g., multiple stiffnesses combinations result in the same observations).

Several papers have illustrated practical engineering examples, in which either the input or response parameters follow multi-modal distributions. In these engineering examples, as either the input or response parameters are shown to follow multi-modal distributions, it is expected that the distribution of the latent uncertain parameters that are affected by either the input or response will also result in a multi-modal distribution. The vibratory load undertaken by the blade of a wind turbine [9], shows a multi-modal distribution when under stochastic excitations. Stresses found at the start and shutdown of generator turbine rotors also show multi-modal distributions [9]. Using long-term monitoring data, it was verified [10,11] that the structural fatigue stress of a steel bridge follows a bi-modal distribution. For nanostructured zirconia coatings, it was observed [12] that the Knoop microhardness follows a bi-modal distribution. Papers [13,14] demonstrated that the axle load spectra can be used for estimating the relative pavement damage of roads. The axle load spectra follow a bimodal distribution as it considers the trucks' weights when unloaded and loaded.

This paper is focused on the inference of multi-modal uncertain parameters of expensive-to-run detailed physics-based models frequently encountered in engineering problems. Within the standard statistical model updating framework [1], the misfit between the features extracted from the measurements and those obtained from the model are used to calculate the likelihood function that is used in the inference scheme. However, because of computational budget and/or time constraints, the number of model evaluations that can be carried out, may be limited. This would significantly hinder the accuracy of the resulting posterior even when applying state-of-the-art statistical model updating approaches. For example, statistical model updating is often implemented by using sampling-based techniques [1]. However, these techniques, including Markov Chain Monte Carlo (MCMC), show a trade-off between computational cost and accuracy, as the convergence of the Markov Chain to the posterior distribution is improved as the chain size lengthens, and therefore the number of physics-based model runs increases [1]. Moreover, Monte Carlo-based techniques introduce a bias, and the number of runs required to achieve convergence is generally unknown when starting the algorithm [1,15].

Alternatively, Variational inference [16] has been used by the machine learning community to estimate posterior distribution approximations employing an optimization approach to reduce the number of model runs required for the inference problem. In simple terms, most variational inference methods propose a family of distributions where the member of the family that best approximates the posterior is chosen [16]. Compared to sampling techniques such as MCMC [17–19], the recent variational inference techniques [20–24] are more numerically scalable and may be used in a wider range of problems due to significant advances in the optimization process [16]. Nevertheless, MCMC based techniques are still the preferred method, as they guarantee convergence to the correct posteriors [23]. However, the disadvantage of these techniques is their high computational cost in order to yield accurate posterior distribution estimates, especially in the presence of highly multi-modal posteriors.

The Variational Bayes Monte Carlo (VBMC) [24,25] has been recently developed to provide an efficient estimation of the model evidence and of the posterior. The method combines active-sampling Bayesian quadrature [26,27] with variational inference [16]. In a nutshell: (a) a postulated posterior is obtained using a Gaussian mixture; (b) the parameters of the Gaussian mixture are obtained using the evidence lower bound (ELBO) as the objective function to be maximised; (c) the expensive to evaluate log unnormalized posterior distribution is replaced by a statistical surrogate model constructed using a GP [28]; (d) active sampling is carried out using 'smart' acquisition functions applied to the GP model to perform a guided local refinement of the GP model; (e) the Bayesian quadrature [26,27] is implemented to carry out fast integrations in the variational objective; (f) a warm-up process is introduced to avoid the algorithm getting initially stuck in areas of very low probability under the true posterior. During the initial phases of the warm-up, significant improvements of the ELBO are rapidly obtained; (g) the algorithm adaptively adjusts the number of components in the variational mixture, adding or removing components based on the level of improvement found on the solution. As a result, the VBMC framework [24,25] is highly efficient. However, the application of VBMC to statistical model updating in engineering problems requires addressing the following challenges: (i) How to select the limited number of initial simulations to build the initial GP? (ii) How to select the new samples to account for multi-modality in the posterior distribution?

To tackle these challenges, the cyclical VBMC approach is proposed. Both first (i) and second (ii) issues, are tackled by introducing an artificial temperature parameter that anneals the unnormalized posterior. This parameter improves the exploration abilities and mode coverage of the algorithm, so the limitations introduced by the limited number of samples and a poor initialization are overcome. This annealing schedule enhances the exploration phase of the cycle and the discovery of regions of high probability density in multi-modal posteriors, as it avoids the algorithm getting stuck in the initially found regions of high probability.

The core of the inference strategy employed in the standard VBMC is the same as the one shown in [29], and it is based on Bayesian quadrature and Variational Inference with a postulated multivariate Gaussian mixture. However, both approaches [29] and [24,25], are such that once a mode of the distribution is found, they are not capable to explore further the uncertain variable domain to identify other modes.

Compared to the work by Ni et al [29], the main differences introduced by the proposed algorithm are: (a) the use of a different acquisition function that selects new points prioritizing the areas of greater probability density compared to the acquisition function based on the absolute value of the mean divided by the standard deviation of the GP surrogate model; (b) variational whitening is performed to deal with posteriors that are highly correlated; (c) convergence criteria based on ELBO compared to the use of criteria related to the vector of variational parameters and the values of the Gaussian mixture weights; (d) the introduction of a warm-up process; (e) the adaptive adjustment of components in the variational mixture; (f) a cyclical annealing schedule to improve the exploration capabilities of the algorithm for dealing with multi-modal posteriors.

The performance of the proposed cyclical VBMC method is assessed with respect to the standard VBMC [24,25], the monotonic VBMC, and the state-of-the-art sampling approach Transitional Ensemble Markov Chain Monte Carlo (TEMCMC) [30] considering both multi-modal and unimodal posteriors of physics-based models parameters.

The paper is structured as follows. In section 2, the Bayesian model updating framework and variational inference are reviewed. The main building blocks on the cyclical VBMC algorithm are described in detail in section 3. The results obtained from three numerical examples and one experimental investigation are presented in section 4. The conclusions of the proposed cyclical VBMC algorithm are then discussed in section 5.

## 2   Bayesian Model Updating Framework

The Bayesian model updating strategy enables the combination of a physics-based model that includes uncertain parameters $\theta$ which cannot be directly observed (also known as latent variables), described by probability density functions, the so-called prior distribution, with new information obtained from measurements of some observable quantities $y_{obs}$ [2,3]. These measurements can be expressed in different forms such as time history, modal properties, etc. This approach results into an updated physics-based model with parameters described as probability density functions, the so-called posterior distributions. This statistical updated model, that is more representative of the real system, can then be used to investigate the behaviour of the system under different loading conditions in order to predict its performance with respect to safety, quality, design or cost constraints [1–4].

In particular, a prior probability density function $p(\theta)$ that reflects the prior knowledge of the uncertain parameters $\theta$ before any measurements on some observable variables $y$ are taken, is assigned to the parameters. Then a likelihood function $p(y_{obs}|\theta)$ that reflects the level of acceptability of the physics-based model, given a set of uncertain parameters $\theta$ to describe the measurements is constructed. This is done by using features extracted (e.g., natural frequencies) from the response obtained with the physics-based model, and the corresponding ones extracted from some measurements. The approach results into an updated statistical physics-based model with its latent variables described as posterior probability density functions $p(\theta|y_{obs})$ that can be calculated using [2,3]:

$$p(\theta|y_{obs}) = \frac{p(\theta)p(y_{obs}|\theta)}{p(y_{obs})} \tag{1}$$

where $p(y_{obs})$ is defined as the evidence, and it serves as a normalization constant for the posterior probability density functions. The posterior $p(\theta|y_{obs})$ can be computed analytically if the prior and likelihood distributions are part of the conjugate family. However, this is not necessarily always the case, and therefore, numerical integration may be necessary.

As the evidence term in Bayesian Inference is a numerical constant, and it is independent of the uncertain parameters $\theta$ [1], sampling techniques (e.g., MCMC) [8,31,32] can be used to obtain samples from the posterior distribution using the following proportional relationship:

$$p(\theta|y_{obs}) \propto p(\theta)p(y_{obs}|\theta) \tag{2}$$

Variational inference takes an alternative approach to sampling methods by minimising the KL divergence between the best member $q(\theta)$ of a postulated family of densities $Q$ and the posterior density, and therefore bypassing the calculation of the evidence term [16]. The posterior distribution is then obtained by transforming the statistical inference problem into an optimization problem. The optimization scheme chooses the member of the family $q(\theta)$ that is 'closest' to the posterior density by converting the minimization of the KL divergence in a maximization of the evidence lower bound (ELBO) objective function. An extensive review of variational inference can be found in [16].

In computer science, several approaches [21–25,33,34] have been developed to circumvent the need of an analytical expression of the ELBO equation, and to reduce the number of evaluations of the physics-based model to be carried out. One of these approaches is the Variational Bayesian Monte Carlo (VBMC) [24,25]. In this paper, a variant of the approach called cyclical VBMC is proposed for addressing the statistical updating problems in engineering where multi-modal posteriors are expected.

## 3 Cyclical Variational Bayesian Monte Carlo framework

The cyclical VBMC approach is based on the VBMC algorithm, it has been developed to deal with multi-modal posteriors in an efficient way by introducing an artificial temperature parameter that anneals the unnormalized posterior. The proposed method overcomes the drawbacks of limited function evaluations and a poor initialization of the VBMC algorithm by introducing the cyclical schedule that improves the exploration abilities and mode coverage of the algorithm.

Given an expensive-to-evaluate computational model of an engineering system, for which prior information on the unknown latent parameters and measurements obtained from the engineering system are available, the proposed approach aims at minimising the number of function evaluations compared to state of the art Bayesian sampling approaches, while obtaining an accurate description of the posterior. The approach consists of two main parts: the initialization of the algorithm and the procedure used to update the parameters in the posterior variational distribution. These two main parts are shown in Fig. 1 and Fig. 2 respectively.

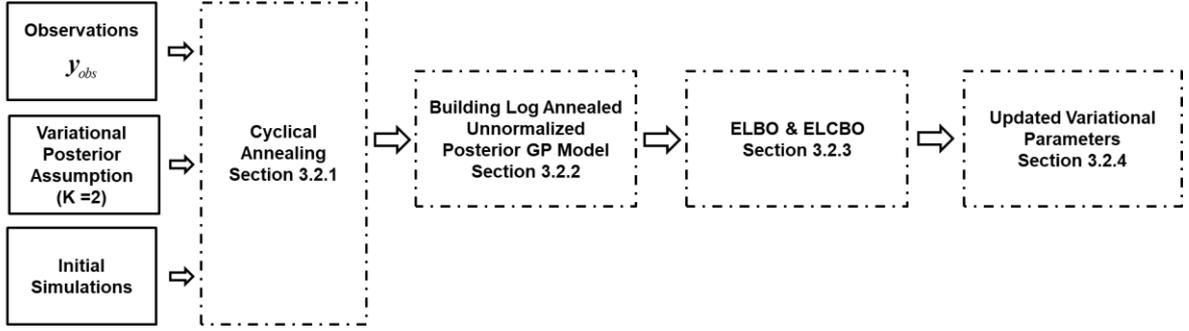

**Fig. 1.** Initialization Blocks of Cyclical VBMC Algorithm.

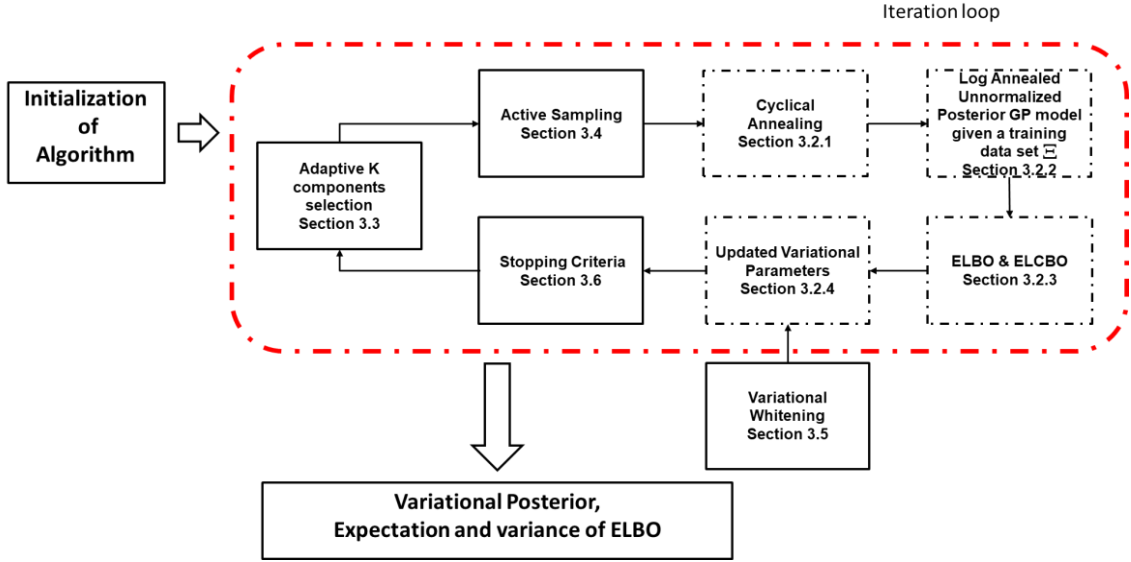

**Fig. 2.** Cyclical VBMC Algorithm Blocks.

The initialization of the algorithm shown in Fig. 1 begins with assuming a variational posterior that is flexible, and able to capture smooth posteriors. This is done by using a Gaussian mixture as the postulated posterior. The multivariate Gaussian mixture model provides a flexible way of describing any continuous density by taking linear combinations of Gaussian distributions (with adjusted means, covariances and the linear combination coefficients). Almost any continuous density can be approximated to arbitrary accuracy [35], therefore, this is chosen as the postulated posterior. Then, an initial set of parameters that is given as input to the physics-based simulation is chosen, and its output is calculated. Given an assumed prior and likelihood function, the logarithm of the unnormalized posterior is calculated. Cyclical annealing is also introduced by replacing the log unnormalized posterior values with the annealed log unnormalized posterior values. A Gaussian Process (GP) regression model using as training points the logarithm of the annealed unnormalized posterior values is employed to build a probabilistic surrogate of the logarithm of the annealed unnormalized posterior. Using Bayesian quadrature [26,27], the GP can then be used to calculate the ELBO and evidence lower confidence bound (ELCBO) values. Finally, the updated variational parameters are obtained.

The second part of the algorithm shown in Fig. 2, consists of a total of $T$ iterations, and it starts with active sampling to select samples for the physics-based model that are run at locations that maximize an acquisition function. The acquisition function is chosen in such a manner that sampling is encouraged at high probability regions of the log unnormalized posterior. Cyclical annealing is introduced into the algorithm by replacing the log unnormalized posterior with the annealed log unnormalized posterior. A prescribed number of cycles and total iterations is set to produce both, an exploration phase and an exploitation phase. The GP regression model is built and the GP's hyperparameters are automatically set by using the maximum-a-posteriori estimates. In this paper it is done by employing initially slice sampling [36], and subsequently gradient based optimization, as recommended to improve computational efficiency in [24,25]. Bayesian Quadrature [26,27] is used to calculate the value of the ELBO. The ELCBO is also calculated, and employed to evaluate the variational approximation's improvement and also as a convergence diagnostic. For each iteration, the stochastic gradient ascent is used to update the parameters of the variational posterior. Variational whitening is also performed every few iterations to deal with highly correlated posteriors. Once the stopping criteria have been fulfilled, the method returns the variational approximation of the posterior, and the expectation and variance of the ELBO.

More details about the approach setup, core building blocks (dashed blocks of Fig. 1 and Fig. 2), adaptive K components selection, active sampling, variational whitening and stopping criteria are given in the following subsections.

## 3.1 Approach Setup

Given a set of observations $y_{obs}$ and a model $PM(x,\theta)$, $x$ represents a vector with fixed model properties known in advance, and $\theta$ is a vector of uncertain model parameters. The setup used in the cyclical VBMC algorithm is described in the following subsections.

### 3.1.1 Variational Approximation of the Posterior $q(\theta)$

The selection of the variational posterior $q(\boldsymbol{\theta})$ is flexible [22], and it should be made with the intent to capture multi-modal posteriors $p(\boldsymbol{\theta}|\mathbf{y}_{obs})$ that can be encountered in engineering applications.

Without loss of generality, the variational posterior $q(\boldsymbol{\theta})$ can be expressed with a non-parametric approximation that is provided by a Gaussian mixture with $K$ components [24,25]:

$$q(\boldsymbol{\theta}) = \sum_{k=1}^{K} w_k \mathcal{N}(\boldsymbol{\theta}; \boldsymbol{\mu}_k, \gamma_k^2 \Sigma) \tag{3}$$

where the mixture weight, mean and scale factor are respectively given by $w_k, \boldsymbol{\mu}_k, \gamma_k^2$, and $\Sigma$ is the covariance matrix:

$$\Sigma = diag\left[\lambda^{(1)^2}, \ldots, \lambda^{(d)^2}\right] \tag{4}$$

with $d$ representing the number of unknown parameters to be inferred (i.e., length of the vector of uncertain parameters $\boldsymbol{\theta}$).

The variational posterior is parameterized in terms of the vector of parameters $\boldsymbol{\phi} \equiv \left(w_1, \ldots, w_K, \boldsymbol{\mu}_1, \ldots, \boldsymbol{\mu}_K, \gamma_1, \ldots, \gamma_K, \lambda\right)$. As a result, the number of parameters to optimise in the variational posterior $q_\phi(\boldsymbol{\theta})$ is given by $d + (d+2)K$, which is the length of the vector $\boldsymbol{\phi}$.

### 3.1.2 Selection of Initial Physics-Based Simulations

Given the uncertain parameters $\boldsymbol{\theta}$, the initial step requires samples to be generated. If available, the plausible lower bound ($PLB$) and the plausible upper bound ($PUB$) which limit the region of the parameter space of high posterior probability mass should be specified. A set of points (as a rule-of-thumb a total of 10 points) situated in the plausible box would be uniformly distributed at random [24,25].

It might occur that the $PLB$ and the $PUB$ are not known. In that case, a set of initial points can be chosen using different sampling methods such as Latin Hypercube Sampling (LHS) [37]. LHS is used in the examples shown in section 4.

The initial points $\boldsymbol{\Theta}_0 = [\boldsymbol{\theta}_1, \ldots, \boldsymbol{\theta}_{n_{init}}]$ are used as inputs for the physics-based model to obtain the output response, where $n_{init}$ is the total number of initial points. The output response can be used to construct features that are then used to evaluate the likelihood values given an assumed likelihood function. The likelihood function reflects the level of agreement between the features obtained by the mathematical model and the measurements. Then, the unnormalized posterior of the initial points can be calculated to build the GP of the annealed logarithm of the unnormalized posterior as described in subsection 3.2.2.

## 3.2 Core Building Blocks

### 3.2.1 Cyclical Annealing

The annealing process is used to flatten the objective function (the ELBO), and to reduce the chance of the algorithm getting stuck in some local optima of the parameters of the variational posterior. The annealing process produces a deterministic deformation of the objective function [38], by means of a temperature parameter [30]:

$$p(\boldsymbol{\theta} | \boldsymbol{y}_{obs})^{\frac{1}{temp}} \propto \left[ p(\boldsymbol{\theta}) p(\boldsymbol{y}_{obs} | \boldsymbol{\theta}) \right]^{\frac{1}{temp}} \tag{5}$$

A fixed temperature implies that the true objective is optimized at a constant schedule, as implemented in the standard VBMC algorithm. Monotonic annealing schedules, in which the temperature is progressively reduced, are the most frequently used. The temperature decreases until the algorithm reaches the true posterior [39].

Although the cyclical annealing schedule [39–42] has been used to deal with multi-modal posteriors in the machine learning field. In this paper, it is the first time that the cyclical annealing schedule is introduced into the combined Variational Inference and Bayesian Quadrature framework (VBMC method) to yield a better representation of the posterior through the introduction of an exploration phase with an improved target guidance. The introduction of the annealing schedule into the VBMC method is of great interest as it is able to remove its inability to deal with multi-modal posteriors. This occurs because the standard VBMC is unable to escape already found regions of high posterior density. The theoretical foundations of the proposed strategy are described in [43].

Specifically, two phases, exploration and exploitation may be found as the temperature is decreased from its maximum to its minimum within each cycle. These phases are cyclically repeated for a prescribed number of times to improve convergence. This enables the algorithm to explore areas of high probability density that may otherwise have not been found. Specifically, during the exploration phase, "paths" would start forming in regions where sampling would take place, producing a high coverage of the support of the target distribution. During the exploitation phase, sampling takes place at regions of high probability density. Therefore, the cyclical schedule gradually improves convergence by reopening paths, and by leveraging on the previous cycles as warm restarts.

The temperature parameter is the inverse of the parameter $\beta_t$:

$$temp = 1/\beta_t \tag{6}$$

The parameter $\beta_t$ is defined in the interval $[0,1]$, and it is calculated for each iteration step in the cyclical VBMC algorithm. According to [39], the $\beta_t$ can be expressed as:

$$\beta_t = \begin{cases} \dfrac{\tau}{S}, & \tau \leq S \\ 1, & \tau > S \end{cases} \tag{7}$$

where:

$$\tau = \frac{\mathrm{mod}(t-1,[S/M])}{S/M} \tag{8}$$

and $t = 1:1:T$ is the iteration number, $T$ is the number of total iterations for the annealing schedule, $M$ is the number of cycles, and $S$ is a control parameter. The exceptional case of $\beta_t = 0$ is circumvented by defining $temp$ as an interval variable $temp \in [1, 50]$. The control parameter $S$ is set to 0.5 as described in [39].

As a rule-of-thumb, if the user has a maximum number of simulations available $N_{sim}$ (e.g., 1000), 20% of these simulations (200) will be allocated for carrying out the cyclical annealing schedule. Therefore, the total number of iterations for the cyclical annealing is obtained by considering the total number of simulations assigned per iteration (for example, 5 simulations per iteration would lead to a total of $T = 40$ iterations of the cyclical annealing schedule). The choice of the number of cycles $M$ depends on the trade-off between exploration and exploitation that the user wants to investigate. For example, if $M = 5$, it would mean that 8 iterations form 1 cycle, in which the temperature (eq.(6)) decreases from its maximum to 1. Once the number of total number of iterations for the cyclical annealing schedule has been reached, the temperature is set to 1.

To introduce the cyclical annealing schedule into the algorithm, the log unnormalized posterior $\log p(\boldsymbol{\theta}, \boldsymbol{y}_{obs})$ is replaced with the annealed log unnormalized posterior $\log p_{annealed}(\boldsymbol{\theta}, \boldsymbol{y}_{obs})$, that is defined as:

$$\log p_{annealed}(\boldsymbol{\theta}, \boldsymbol{y}_{obs}) = \frac{1}{temp} \log p(\boldsymbol{\theta}, \boldsymbol{y}_{obs}) \tag{9}$$

### 3.2.2 Gaussian Process (GP) of the Annealed Logarithm Unnormalized Posterior

For the proposed cyclical VBMC, cyclical tempering is introduced into the algorithm, by simply replacing the log unnormalized posterior $f \equiv \log p(\boldsymbol{\theta}, \boldsymbol{y}_{obs})$ with the annealed log unnormalized posterior $f \equiv \log p_{annealed}(\boldsymbol{\theta}, \boldsymbol{y}_{obs})$.

The annealed log unnormalized posterior $f \equiv \log p_{annealed}(\boldsymbol{\theta}, \boldsymbol{y}_{obs})$, is approximated using a GP regression model [28]:

$$f \sim GP\big(m_{gp}(\boldsymbol{\theta}), k_{gp}(\boldsymbol{\theta}, \boldsymbol{\theta}')\big) \tag{10}$$

where $m_{gp}(\boldsymbol{\theta})$ is the mean function, and a covariance matrix is defined in terms of a kernel function $k_{gp}(\boldsymbol{\theta}, \boldsymbol{\theta}')$. The typical choice when little is known about the function to be approximated [28] is to use the squared exponential kernel that is expressed as:

$$k_{gp}(\boldsymbol{\theta}, \boldsymbol{\theta}') = \sigma_f^2 \Lambda \exp\left(-\frac{1}{2}(\boldsymbol{\theta}-\boldsymbol{\theta}')^T \Sigma_l^{-1}(\boldsymbol{\theta}-\boldsymbol{\theta}')\right) \tag{11}$$

Where $\sigma_f$ is the output length scale, and:

$$\Lambda \equiv (2\pi)^{\frac{d}{2}} \prod_{i=1}^{d} l^{(i)} \tag{12}$$

is the normalization of the Gaussian, $\boldsymbol{l}$ is the vector of input length scales, the superscript $^{(i)}$ refers to the $i$-th dimension and $\Sigma_l$ is a diagonal covariance matrix:

$$\Sigma_l = diag(l^{(i)^2}, \ldots, l^{(d)^2}) \tag{13}$$

The likelihood is assumed to be Gaussian with an observation noise $\sigma_{obs} > 0$ (to obtain numerical stability [44]), and the shape of the mean function is assumed to be given by [24,25]:

$$m_{gp}(\boldsymbol{\theta}) = m_0 - \frac{1}{2}\sum_{i=1}^{d}\frac{\left(\theta^{(i)} - \theta_{max}^{(i)}\right)^2}{r^{(i)^2}} \tag{14}$$

Where $m_0$ is the mean's maximum value, $\boldsymbol{\theta}_{max}$ is the location of the mean's maximum value, and $\boldsymbol{r}$ is the length scales' vector [24,25].

The hyperparameters that define the GP are collected in a vector $\boldsymbol{\psi} = \left[\boldsymbol{l}, \sigma_f, \sigma_{obs}, m_0, \boldsymbol{\theta}_m, \boldsymbol{r}\right]$, of dimension $3d+3$. These hyperparameters $\boldsymbol{\psi}$, are themselves defined in terms of a uniform distribution or a truncated Student's $t$ distribution with mean $\mu$, standard deviation $\sigma$, and $\nu = 3$ degrees of freedom. Some of these GP hyperparameters $\left[\boldsymbol{l}, \sigma_f, \sigma_{obs}, \boldsymbol{r}\right]$ are defined in the log space. The same distributions used in references [25] have been directly implemented and are given in Table 1.

**Table 1**

Table of hyperparameters' prior.

| Hyperparameter | Description | Prior mean | Prior scale |
|---|---|---|---|
| $\log l^{(i)}$ | Input length scale ($i$-th dimension) | $\log\left[\sqrt{\frac{d}{6}}L^{(i)}\right]$ | $\log\sqrt{10^3}$ |
| $\log \sigma_f$ | Output scale | Uniform | - |
| $\log \bar{\sigma}_{obs}$ | Base observation noise | $\log\sqrt{10^{-5}}$ | 0.5 |
| $m_0$ | Mean function maximum | Uniform | - |
| $\theta_{max}^{(i)}$ | Mean function location ($i$-th dimension) | Uniform | - |
| $\log r^{(i)}$ | Mean function scale ($i$-th dimension) | Uniform | - |

$L^{(i)}$ is defined by:

$$L^{(i)} = PUB^{(i)} - PLB^{(i)} \tag{15}$$

Where the *PLB* and *PUB* limit the region of the parameter space of high posterior probability mass.

The calculation of the maximum-a-posteriori (MAP) of the hyperparameters may be performed by any appropriate algorithm. Another possible option to calculate the hyperparameters would be to use the maximum-likelihood-estimate (MLE) method. In this paper, the MAP of the hyperparameters is first calculated using slice sampling, and the estimation is subsequently switched to a gradient based optimization approach when the variability of the expected log unnormalized posterior is below a threshold [24,25].

By conditioning, the resulting GP predictive posterior mean function $\overline{f}_\Xi(\theta)$ and posterior covariance function $C_\Xi(\theta, \theta')$ for a training data set $\Xi = \{\Theta, h, \sigma_{obs}\}$ (for N training inputs $\Theta = [\theta_1, \ldots, \theta_N]$, and $N$ observations $h = f(\Theta)$ with observation noise $\sigma_{obs} > 0$) is given in closed-form [28] as:

$$\overline{f}_\Xi(\theta) \equiv \mathbb{E}\left[f(\theta) \mid \Xi, \psi\right] = k(\theta, \Theta)\left[k(\Theta, \Theta) + \Sigma_{obs}(\Theta)\right]^{-1}(h - m(\Theta)) + m(\Theta) \tag{16}$$

$$C_\Xi(\theta, \theta') \equiv Cov\left[f(\theta), f(\theta') \mid \Xi, \psi\right] = k(\theta, \theta') - k(\theta, \Theta)\left[k(\Theta, \Theta) + \Sigma_{obs}(\Theta)\right]^{-1} k(\Theta, \theta')$$

(17)

The observation noise matrix has the following form:

$$\Sigma_{obs} = diag(\sigma_{obs}^2(\theta_1), \ldots, \sigma_{obs}^2(\theta_N)) \tag{18}$$

As the annealed log unnormalized posterior $\log p_{annealed}(\theta, y_{obs})$ is approximated using a GP model, an analytical computation of the integral involved in the ELBO and ELCBO equation can be derived using Bayesian Quadrature [26,27], as described in what follows.

### 3.2.3 The Evidence Lower Bound (ELBO) and Evidence Confidence Lower Bound (ELCBO)

The ELBO can now be expressed as [24]:

$$ELBO(\phi, f) = \mathbb{E}_q\left[\log p(\theta, y_{obs})\right] - \mathbb{E}_q\left[\log q(\theta)\right] = \mathbb{E}_{f|\Xi}\left[\mathbb{E}_\phi[f]\right] + \mathcal{H}[q(\theta)] \tag{19}$$

Where $\mathcal{H}[q(\boldsymbol{\theta})]$ is the entropy of the variational posterior [24]. The integrals in eq.(19) can be analytically computed [24,25] with the Bayesian MC statistical method also known as Bayesian quadrature, so that [26,27]:

$$\mathrm{E}_{f|\Xi}[Z] = \int \bar{f}_{\Xi}(\boldsymbol{\theta})\pi(\boldsymbol{\theta})d\boldsymbol{\theta} \qquad (20)$$

$$\mathrm{V}_{f|\Xi}[Z] = \iint C_{\Xi}(\boldsymbol{\theta},\boldsymbol{\theta}')\pi(\boldsymbol{\theta})\pi(\boldsymbol{\theta}')d\boldsymbol{\theta}d\boldsymbol{\theta}' \qquad (21)$$

where:

$$Z = \int_{\mathrm{R}^d} g(\boldsymbol{\theta})\pi(\boldsymbol{\theta})d\boldsymbol{\theta} \qquad (22)$$

In the proposed cyclical VBMC algorithm the function $g(\boldsymbol{\theta})$ is given by the annealed log unnormalized posterior $\log f(\boldsymbol{\theta}) \equiv \log p_{annealed}(\boldsymbol{\theta}, \boldsymbol{y}_{obs})$ and $\pi(\boldsymbol{\theta})$ is the variational approximation to the posterior $q(\boldsymbol{\theta})$ [11,13].

The variational approximation's $\mathcal{H}[q(\boldsymbol{\theta})]$ entropy is calculated using Monte Carlo sampling, and the gradient is propagated using a reparameterization trick [34,45], which allows stochastic gradient ascent [46] to be used to optimize the ELBO equation.
The evidence lower confidence bound (ELCBO) is [24]:

$$ELCBO(\boldsymbol{\phi}, f) = \mathrm{E}_{f|\Xi}\left[\mathrm{E}_{\phi}[f]\right] + \mathcal{H}[q(\boldsymbol{\theta})] - \beta_{LCB}\sqrt{\mathrm{V}_{f|\Xi}\left[\mathrm{E}_{\phi}[f]\right]} \qquad (23)$$

where the term $\beta_{LCB}$ represents a risk-sensitivity term.
The ELCBO is the probabilistic lower bound of the ELBO, and it can be used to judge the variational approximation's improvement. As $\mathrm{V}_{f|\Xi}\left[\mathrm{E}_{\phi}[f]\right]$ in eq.(23) decreases, the ELCBO value will converge to the ELBO value [24].
The first two terms of the ELCBO equation are estimated as described before. The risk sensitivity term [24] is usually set to $\beta_{LCB} = 3$.

### 3.2.4 Update of Variational Parameters

The variational posterior is parameterized in terms of the variational parameters in the vector $\boldsymbol{\phi}$. These variational parameters are updated by solving an optimization problem [24,25]:

$$\hat{\boldsymbol{\phi}} = \arg\max_{\boldsymbol{\phi}} \{ELBO(\boldsymbol{\phi}, f)\} \qquad (24)$$

This optimization is carried out using a stochastic gradient descent algorithm based on a variant of Adam [47] to obtain the updated variational posterior.

### 3.3 Adaptive K Components Selection

A warm-up stage is used in the initial iterations of the algorithm. In this phase, the variational posterior is specified in terms of a $K = 2$ Gaussian mixture with $w_1 \equiv w_2 = 0.5$. The warm-up phase finishes when the improvement of the ELCBO for three consecutive iterations is smaller than 1, this implies that the variational solution is stabilizing.

An adequate number of components $K$ should be used to capture the true posterior [22]. The number of components in the Gaussian mixture used for the approximation is adaptively chosen as described in [24]. For this purpose, a component can be added to or removed from the Gaussian mixture. A component is added to the Gaussian mixture after the ELCBO of the current iteration is greater than the ELCBO found in the last four iterations. This is done as long as during the last iteration no mixture component was removed. An additional condition can also be set to speed up the approximation as explained in [24,25]. A component of the Gaussian mixture can also be randomly removed from the variational solution, if it simultaneously occurs that the mixture weight is smaller than 0.01, and the difference between the ELCBO of the variational solution in this iteration, and the ELCBO of the variational solution after removal of that component, is smaller than 0.01. More information can be found in [24,25].

### 3.4 Active Sampling

Active sampling is employed to select a number of prescribed samples within each iteration at locations which maximize an acquisition function.

These samples are the input parameters for which the physics-based model is evaluated. The acquisition function $a_{pro}$, is chosen in such a manner that sampling is encouraged at high probability regions of the log unnormalized posterior [24,25]:

$$a_{pro}(\boldsymbol{\theta}) = s_{\Xi}^2(\boldsymbol{\theta}) \exp(f_{\Xi}(\boldsymbol{\theta})) q_{\phi}(\boldsymbol{\theta}) \tag{25}$$

where $s_{\Xi}^2(\boldsymbol{\theta})$ is the variance of the GP posterior, $\exp(f_{\Xi}(\boldsymbol{\theta}))$ is the exponentiated GP posterior mean for a given training set $\Xi$, and $q_{\phi}(\boldsymbol{\theta})$ is the variational approximation of the posterior at $\boldsymbol{\theta}$.

Therefore, this acquisition function favours mostly exploitation of the knowledge obtained in the previous iterations. To add an exploration phase, cyclical annealing is also introduced. This enables the algorithm to explore areas of high probability density that may otherwise have not been found. As shown in the first and second numerical example, cyclical annealing allows sampling of multi-modal regions as an exploration phase occurs in the initial iterations of each annealing cycle. Due to the exploitation nature of the acquisition function, the last iterations of each annealing cycle start sampling at the already found modes.

### 3.5 Variational Whitening

To deal with highly correlated posterior distributions, variational whitening is introduced in the cyclical VBMC algorithm. This is carried out using a linear transformation of the inference space to a new space where the covariance matrix results in a unit diagonal matrix [25]. The transformation matrix $W$ (rotation and scaling) is obtained using singular value decomposition (SVD) of the covariance matrix of the variational posterior $q(\boldsymbol{\theta})$. Variational whitening occurs after the reliability index $\rho(t)$ described in the next subsection is lower and or equal to 3. It is applied in increasingly spaced intervals within iterations as illustrated in [25].

### 3.6 Stopping Criteria

To determine the number of required iterations, the algorithm uses a reliability index $\rho(t) \geq 0$, that suggests the stability of the variational solution. The algorithm is finished if the value $\rho(t) \leq 1$ is found at the end of $n_{stable}$ = 8 consecutive iterations, where a maximum of one intermediate iteration may be unstable, or if a predetermined number of iterations $n_{max}$ is reached.

The value at iteration $t$ of the reliability index is calculated as the average of the three reliability features $\rho_j(t)$ for $j = 1, 2, 3$:

$$\rho(t) = \frac{\rho_1(t) + \rho_2(t) + \rho_3(t)}{3} \tag{26}$$

The value of the reliability index $\rho_1(t)$ is calculated as a function of the KL divergence between the previous and the current variational posterior. The reliability index $\rho_2(t)$ is a function of the change of ELBO between two consecutive iterations, and $\rho_3(t)$ is a function of the estimation of the variance of the ELBO. These indices give an overall measure of how the variational posterior is converging throughout the iterations and are defined as [24,25]:

$$\rho_1(t) = \frac{\left| E\left[ ELBO(t) \right] - E\left[ ELBO(t-1) \right] \right|}{\Delta_{SD}} \tag{27}$$

$$\rho_2(t) = \frac{\sqrt{V\left[ ELBO(t) \right]}}{\Delta_{SD}} \tag{28}$$

$$\rho_3(t) = \frac{KL(q_t \| q_{t-1}) + KL(q_{t-1} \| q_t)}{2\Delta_{KL}} \tag{29}$$

The parameters $\Delta_{KL}$ and $\Delta_{SD}$ should be chosen in such a manner that the values of the individual reliability features meet the inequality $\rho_j \lesssim 1$, where $j = 1, 2, 3$, for the values of $\rho(t)$ considered representative of good solutions. In the cyclical VBMC algorithm, the values of $\Delta_{SD}$ and $\Delta_{KL}$ are respectively set at 0.1 and $0.01\sqrt{d}$.

### 3.7 Steps of the Approach

The proposed approach can be summarised in the following steps:

1. Initialization of Algorithm (Fig. 1):
    a. Initial training set for physics-based simulation is run.
    b. Cyclical Annealing is introduced (eq.(9)).
    c. Logarithm of (annealed) unnormalized posterior of the initial set is calculated.
    d. GP surrogate model of the (annealed) logarithm unnormalized is built using the initial training set values calculated in 1.b.
    e. ELBO and ELCBO are calculated (eq.(19) and eq.(23)).
2. Second part of the algorithm (Fig. 2):
    a. Selection of new samples using an acquisition function (eq.(25)), these are used to actively update the GP surrogate model.
    b. Cyclical Annealing is introduced (eq.(9)).
    c. The GP surrogate model of the (annealed) logarithm unnormalized posterior is built.
    d. Calculation of ELBO and ELCBO value (eq.(19)) and eq.(23)).

e. Update the variational parameters (variational whitening may also be applied at this step).
   f. Check if stopping criteria have been met, if not repeat from step 2.a.

Where the main outputs of the algorithm are the variational posterior, the expected value of the ELBO, and the variance of the ELBO.

## 4 Results Obtained from the Case Studies Investigated

This section has the purpose of comparing the proposed cyclical VBMC algorithm with the standard VBMC algorithm, the monotonic VBMC algorithm and the Transitional Ensemble Markov Chain Monte Carlo (TEMCMC) sampling algorithm. The functions `plotmatrix` and `ksdensity` from MATLAB [48] were used to plot the posterior distributions obtained with the TEMCMC algorithm.

Four case studies have been chosen to showcase the advantages and disadvantages of the cyclical VBMC for different distribution modality: one highly multi-modal (first example), one mildly multi-modal (second example), one unimodal (third example) numerical examples and a multi-modal experimental case study.

The number of function evaluations (number of samples), the number of iterations used to achieve convergence (this will determine the number of function evaluations needed), and the number of modes found, are used to compare the performance of the algorithms on the multi-modal examples. On the unimodal example, for the same purpose, the samples mean, coefficient of variation, number of function evaluations and number of iterations are used. For all the VBMC implementations, 300,000 samples of the variational posterior are taken to compute the sample mean and sample coefficient of variation.

Ten initial samples are picked using LHS [37] for all the case studies that use a form of the VBMC algorithm, and for every iteration that occurs within the algorithm, five samples are chosen using the acquisition function, and are evaluated. The samples chosen, correspond to evaluations of the physics-based model.

The monotonic annealing schedule used in the monotonic VBMC is calculated for a total number of iterations $T = 40$, with one cycle $M = 1$ and a parameter $S = 0.5$. However, for the cyclical annealing schedule in the cyclical VBMC, the number of cycles is $M = 5$. The monotonic schedule maximum temperature of fifty is subsequently decreased in each iteration until a minimum temperature of one is reached. The same concept is applied to the cyclical annealing schedule that has five cycles where a pre-set maximum temperature of fifty is subsequently decreased in each iteration until a minimum temperature of one is reached in each cycle.

Throughout the examples it will be shown that the monotonic VBMC and cyclical VBMC require a higher amount of samples evaluations compared to the standard VBMC for problems with low dimensionality (i.e., low number of inferred parameters). This is expected as a total number of iterations $T = 40$ is prescribed for both algorithms, meaning that forty is the lowest number of iterations possible.

### 4.1 Himmelblau Multi-modal Posterior

A multi-modal problem based on [30,49] is introduced in this subsection. The posterior used (4 peaks, 2-dimensional) can be observed on Fig. 3 and it has as its mathematical expression [49] the Himmelblau's function $HB(\theta_1, \theta_2)$:

$$HB(\theta_1, \theta_2) = \left(\theta_1^2 + \theta_2 - 11\right)^2 + \left(\theta_1 + \theta_2^2 - 7\right)^2 \tag{30}$$

The $HB(\theta_1, \theta_2)$ is frequently used to assess the performance of optimization algorithms. It has four distinct solutions of local minima at $\{\theta_1, \theta_2\}_1 = \{3, 2\}$, $\{\theta_1, \theta_2\}_2 = \{-2.805, 3.131\}$, $\{\theta_1, \theta_2\}_3 = \{-3.779, -3.283\}$ and $\{\theta_1, \theta_2\}_4 = \{-3.584, -1.848\}$.

The posterior of interest is then defined as follows [30,49]:

$$p(\boldsymbol{\theta} | \boldsymbol{y}_{obs}) \propto \exp\left[-HB(\theta_1, \theta_2)\right] \tag{31}$$

That ensures that the local minima of $HB(\theta_1, \theta_2)$ become regions of high probability density, producing the 4 peaks shown in Fig. 3. The likelihood function is modelled as the exponential function of $-HB(\theta_1, \theta_2)$, and thus takes the same mathematical form as the posterior [30,49]. The uniform priors $\theta_1 \sim U(-5, 5)$ and $\theta_2 \sim U(-5, 5)$ have been used.

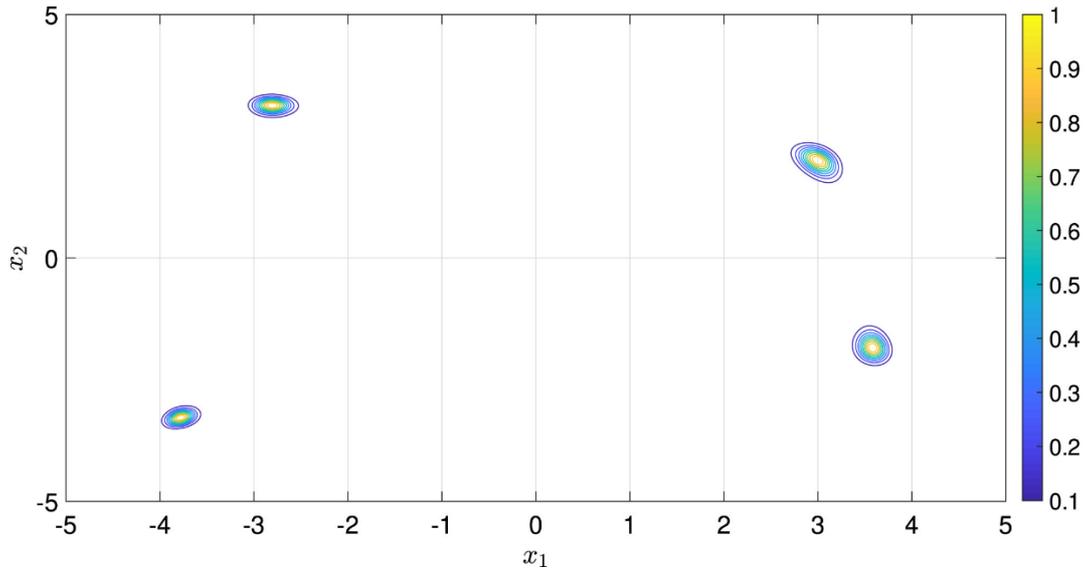

**Fig. 3.** Contour plot of the Himmelblau's function using eq.(31) taken from [30]. The values of the posterior are given by the numbers on the colour chart.

The results found using the standard VBMC algorithm after running several iterations are illustrated in Fig. 4. The final 1-D and 2-D marginal posterior distributions obtained by the standard VBMC algorithm are shown in Fig. 5. The figures show that only one mode has been found, as due to the nature of the algorithm, the active sampling used is unable to escape from that mode. In other words, the algorithm proceeds to only sample in the vicinity of that mode due to its exploitation nature. A total of 75 function evaluations were required, and only one mode was found.

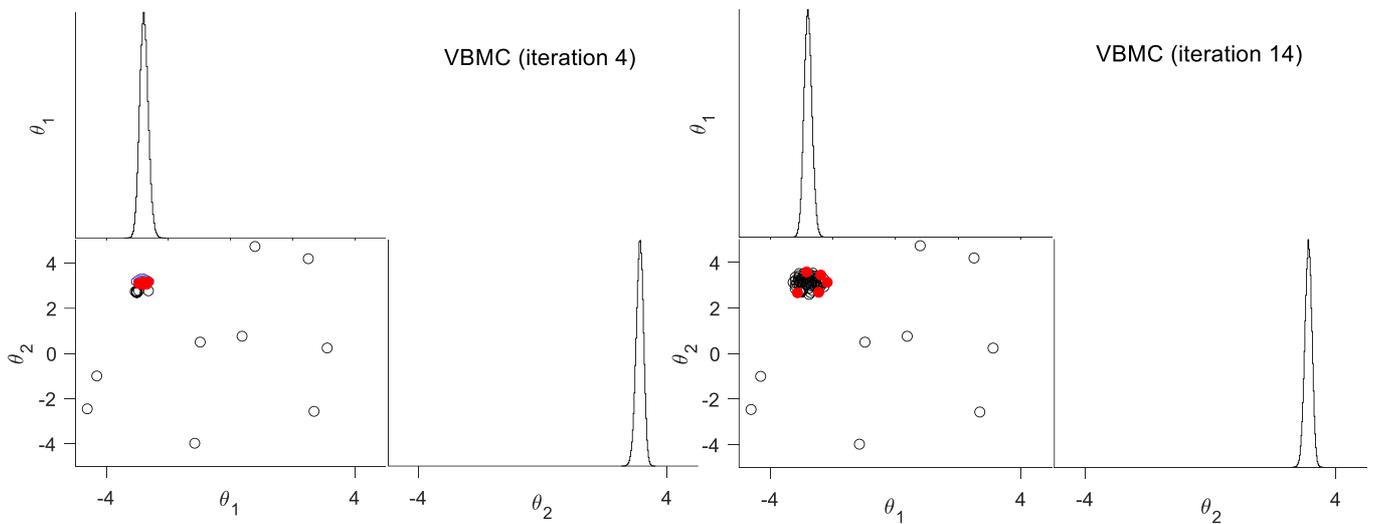

**Fig. 4.** Resulting 1-D and 2-D marginal posterior distributions from VBMC at 4$^{th}$ and 14$^{th}$ iterations. Red dots indicate samples taken at the current iteration. Black circles indicate samples used for the GP model of the unnormalized posterior at each iteration.

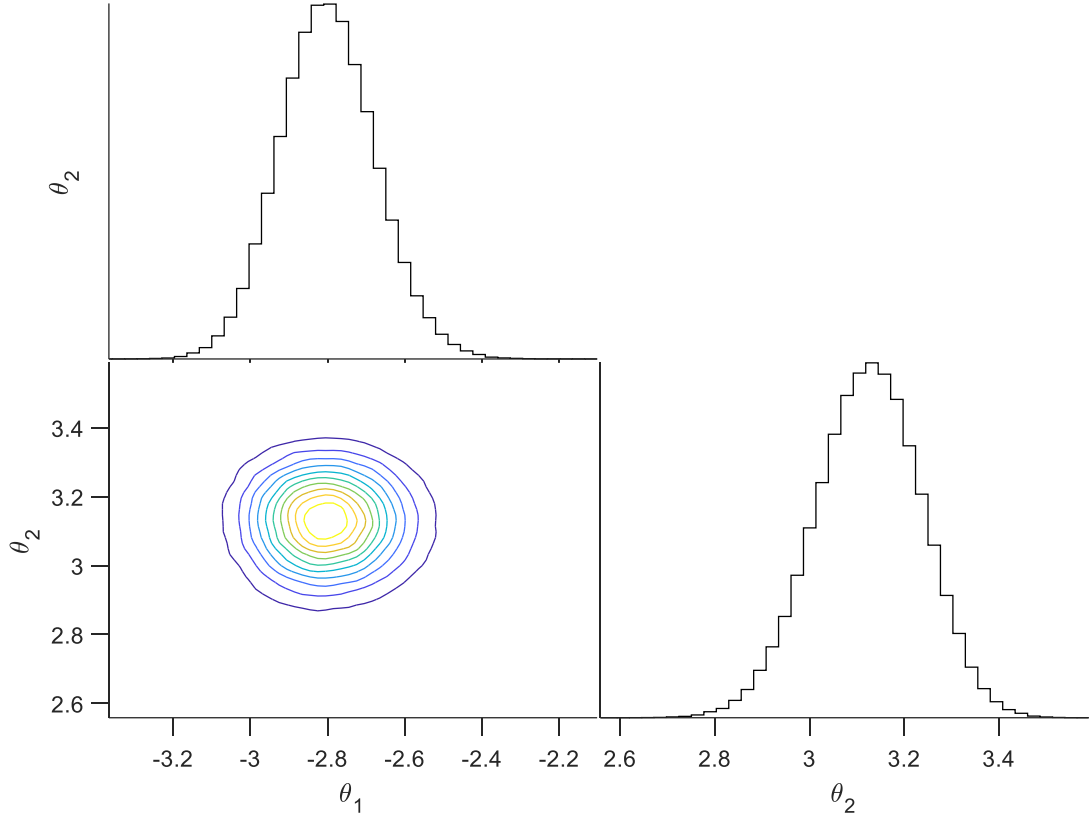

**Fig. 5.** Final 1-D and 2-D marginal posterior distributions from VBMC algorithm.

The results obtained with the monotonic and the cyclical VBMC algorithm are shown in Fig. 6 and Fig. 7, it is found that the overall results of these two schedules significantly differ.

It can be seen in Fig. 6 that in the monotonic VBMC algorithm, for the first few iterations, the samples are chosen following an exploration approach. In the final iterations of the monotonic algorithm shown in Fig. 7, the samples chosen are close to the two modes found. The resulting refined postulated posterior of the algorithm when using a monotonic annealing schedule is only able to account for two modes. The monotonic VBMC algorithm needs a total of 235 function evaluations to converge to the 2-mode estimated posterior shown in Fig. 8.

It can be seen in Fig. 6, that in the proposed cyclical VBMC algorithm, for the first few iterations, the samples are chosen following both an exploration and an exploitation approach. For the final iterations of the cyclical algorithm, it can be seen in Fig. 7, that the samples chosen are close to the four modes found. The resulting refined postulated posterior of the algorithm, using the cyclical annealing schedule shown in Fig. 8, is able to account for all four modes. A total of 300 function evaluations were needed to converge to the estimated posterior.

The empirical cumulative density functions (ECDFs), shown in Fig. 9 were calculated using samples obtained from both the cyclical VBMC, and the TEMCMC algorithm, applying the function `cdfplot` from Matlab [48]. For the TEMCMC algorithm, the number of samples used to calculate the ECDFs were progressively increased until convergence occurred. It was found that the converged ECDFs obtained by the TEMCMC algorithm, required a much larger number of samples than the cyclical VBMC algorithm.

The computational cost is significantly reduced for the three VBMC algorithms compared to the TEMCMC sampling algorithm, where 5000 evaluations of the likelihood function were needed as shown in [30]. The proposed cyclical VBMC algorithm is the only VBMC variant that is able to find the four modes of the Himmelblau posterior. The numerical results for the Himmelblau multi-modal posterior are summarised in Table 2.

**Table 2**

Comparison of numerical results for the Himmelblau multi-modal posterior.

| Method | N. of samples | N. of Total Iterations for Convergence | N. of modes found |
|---|---|---|---|
| VBMC | 75 | 14 | 1 |
| Monotonic VBMC | 235 | 46 | 2 |
| Cyclical VBMC | 300 | 59 | 4 |
| TEMCMC [37] | 5000 | 5 | 4 |

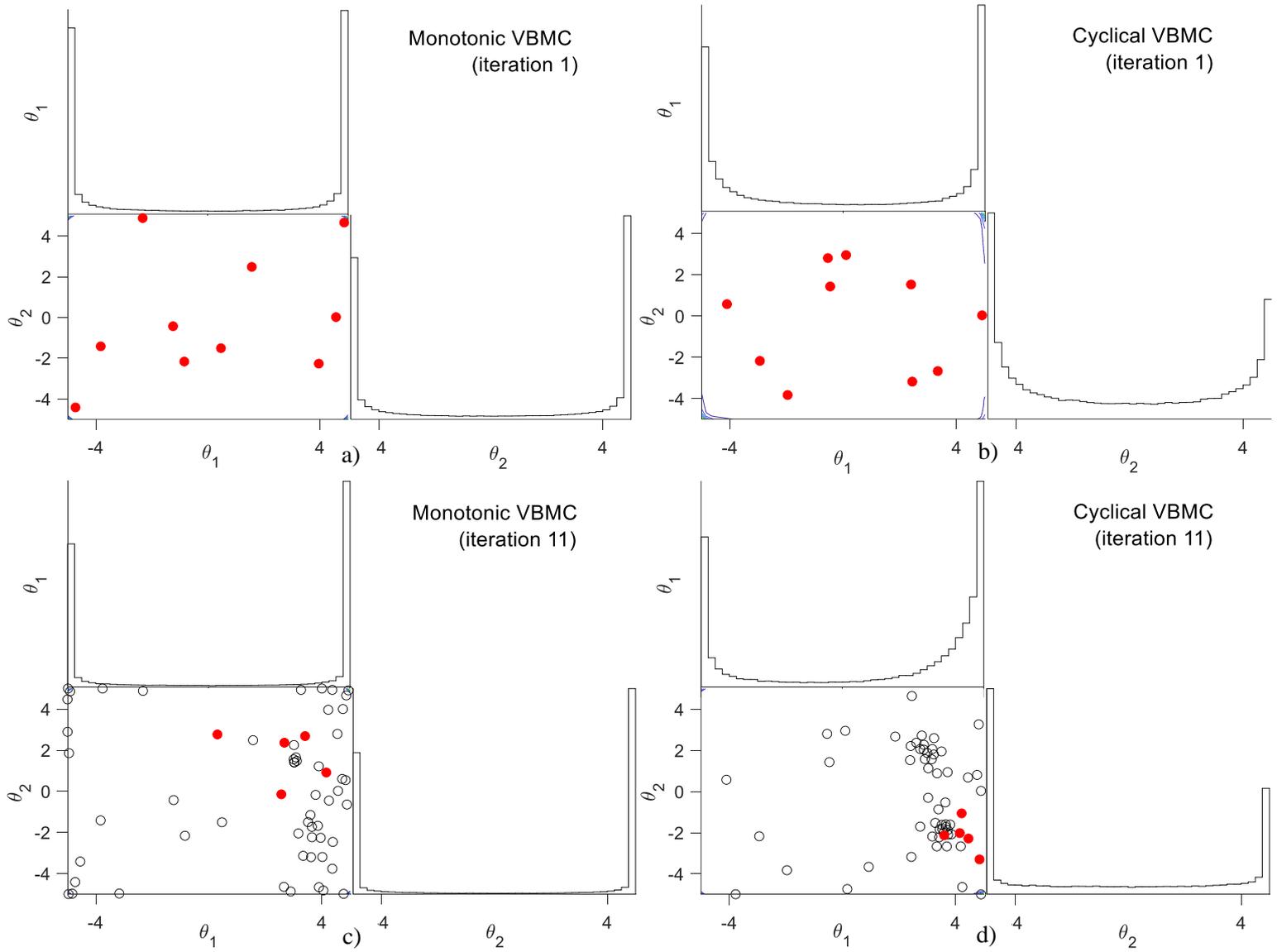

**Fig. 6.** Resulting 1-D and 2-D marginal posterior distributions from VBMC at 1st and 11th iterations. Red dots indicate samples taken at the current iteration. Black circles indicate samples used for the GP model of the unnormalized posterior at each iteration. a) Monotonic annealing schedule at 1st iteration; b) Cyclical Annealing Schedule at 1st iteration; c) Monotonic annealing schedule at 11th iteration; d) Cyclical Annealing Schedule at 11th iteration.

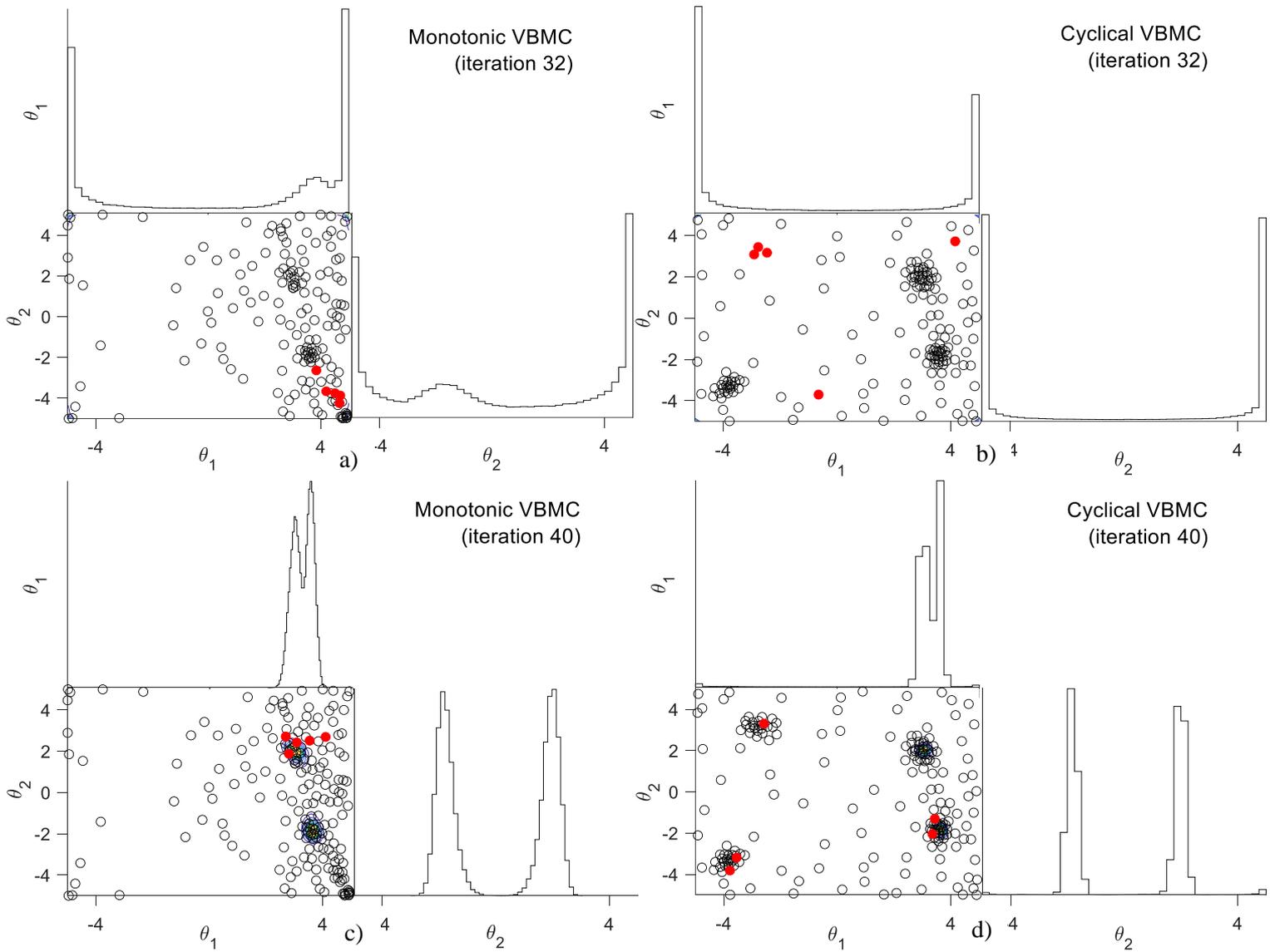

**Fig. 7.** Resulting 1-D and 2-D marginal posterior distributions from VBMC at 32$^{nd}$ and 40$^{th}$ iterations. Red dots indicate samples taken at the current iteration. Black circles indicate samples used for the GP model of the unnormalized posterior at each iteration. a) Monotonic annealing schedule at 32$^{nd}$ iteration; b) Cyclical Annealing Schedule at 32$^{nd}$ iteration; c) Monotonic annealing schedule at 40$^{th}$ iteration; d) Cyclical Annealing Schedule at 40$^{th}$ iteration.

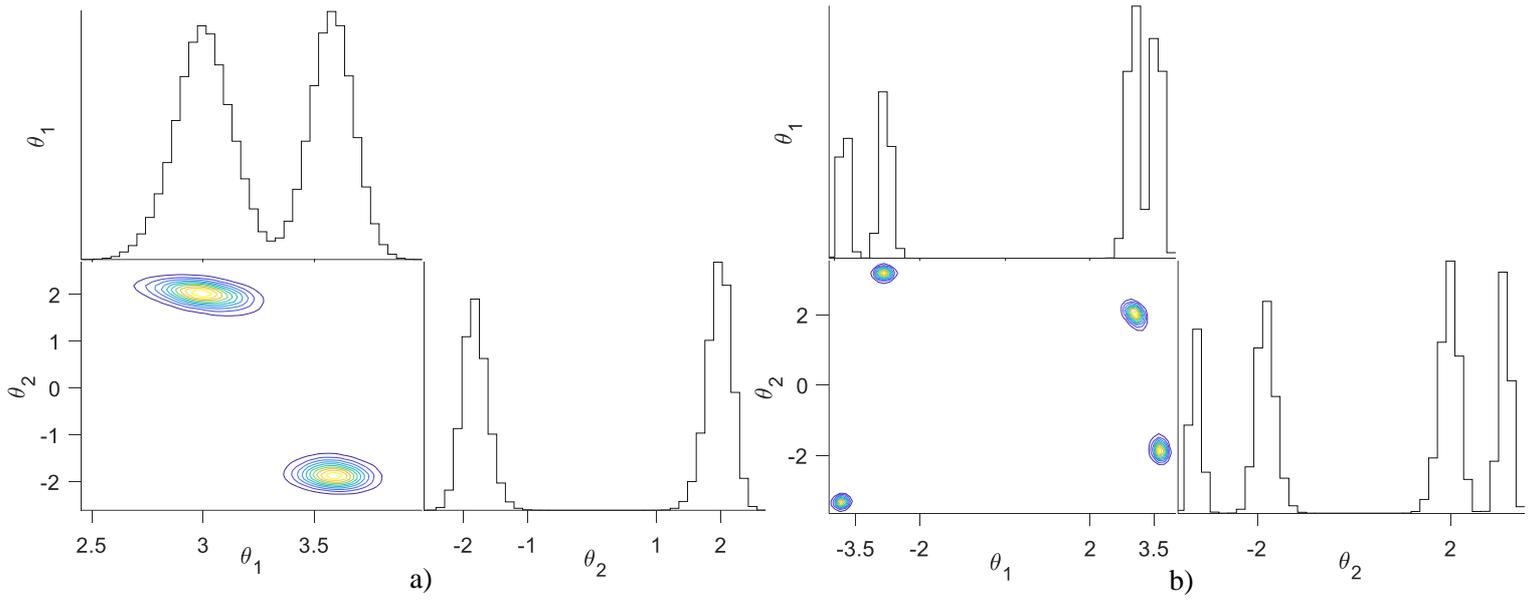

**Fig. 8.** Final 1-D and 2-D marginal posterior distributions from VBMC algorithm. a) Monotonic annealing schedule; b) Cyclical Annealing Schedule.

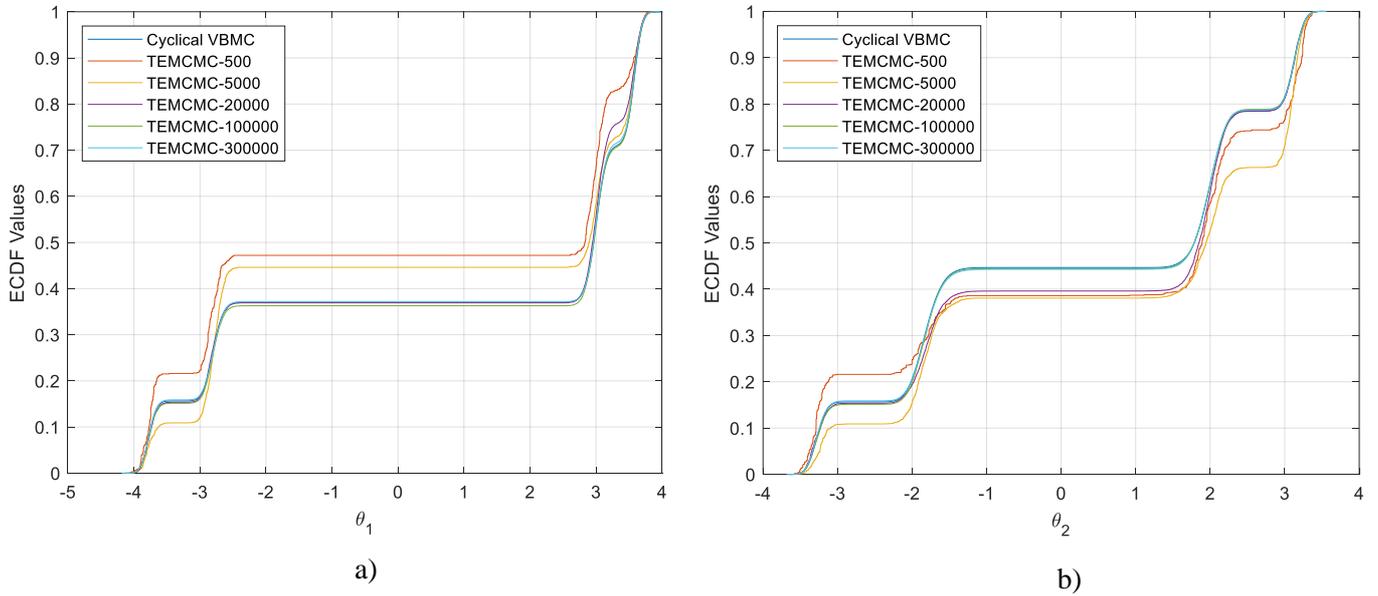

**Fig. 9.** Marginal ECDFs using Cyclical VBMC and TEMCMC. a) Parameter $\theta_1$; b) Parameter $\theta_2$.

## 4.2 Mass-spring System (Multi-modal Posterior)

In this example taken from [50], a 2-dimensional multi-modal Bayesian model updating system that may be found in engineering problems, is used to compare the differences in the performance of the aforementioned algorithms. For the purposes of this subsection, the numerical performance will be based on the number of samples used to compute the posterior, the number of modes found in each algorithm and the empirical cumulative density function.

As shown in Fig. 10, a 2-degrees of freedom (2-DoF) system with masses $m_1 = 16.531 \times 10^3 kg$, $m_2 = 16.131 \times 10^3 kg$ joined by springs with stiffness $k_1 = \bar{k}\theta_1$, $k_2 = \bar{k}\theta_2$, where $\bar{k} = 29.7 \times 10^6 N/m$ is defined, and $\theta_1$ and $\theta_2$ are the uncertain parameters to be inferred. For the spring constants, the uniform priors $\theta_1 \sim U(0.01, 3)$ and $\theta_2 \sim U(0.01, 3)$ have been used.

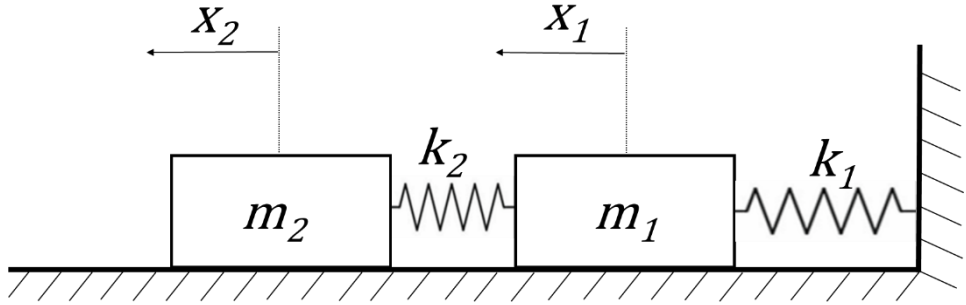

**Fig. 10.** First mass-spring system.

Two independent likelihood functions are used with standard deviations $\sigma_i = 0.02\omega_i$ (2% of the deterministic values of the natural frequencies) and with means that equal the deterministic values of the natural frequencies.

From Fig. 11, it can be observed that the standard VBMC is only able to find one mode, and that the active sampling is again unable to escape from that mode. The standard algorithm uses a total of 75 function evaluations to obtain the resulting posterior distribution that only accounts for a single mode. The final 1-D and 2-D marginal posterior distributions obtained by the standard VBMC algorithm are shown in Fig. 12.

For the monotonic VBMC algorithm, an exploration phase is shown in the first few iterations. An exploitation phase that samples in the vicinity of the two found modes is illustrated in Fig. 13.

However, for the cyclical VBMC algorithm both exploration and exploitation occur in the early iterations, as shown in Fig. 13. The final 1-D and 2-D marginal posterior distributions obtained by the monotonic and cyclical VBMC algorithm are shown in Fig. 14, where it is possible to observe that both methods account for the two modes.

In Fig. 16 the ECDFs for the cyclical VBMC, monotonic VBMC, and TEMCMC are shown. The resulting ECDFs are found to be similar, with a slight difference observed for the ECDFs obtained from the TEMCMC algorithm.

Table 3 summarises the number of evaluations and iterations needed for the three analysed VBMC algorithms, and the sampling method TEMCMC (Fig. 15). It can be seen that when using the three VBMC algorithms, the computational cost is significantly reduced compared to the TEMCMC sampling algorithm, where 5000 evaluations of the likelihood function were needed to obtain samples from the posterior distribution.

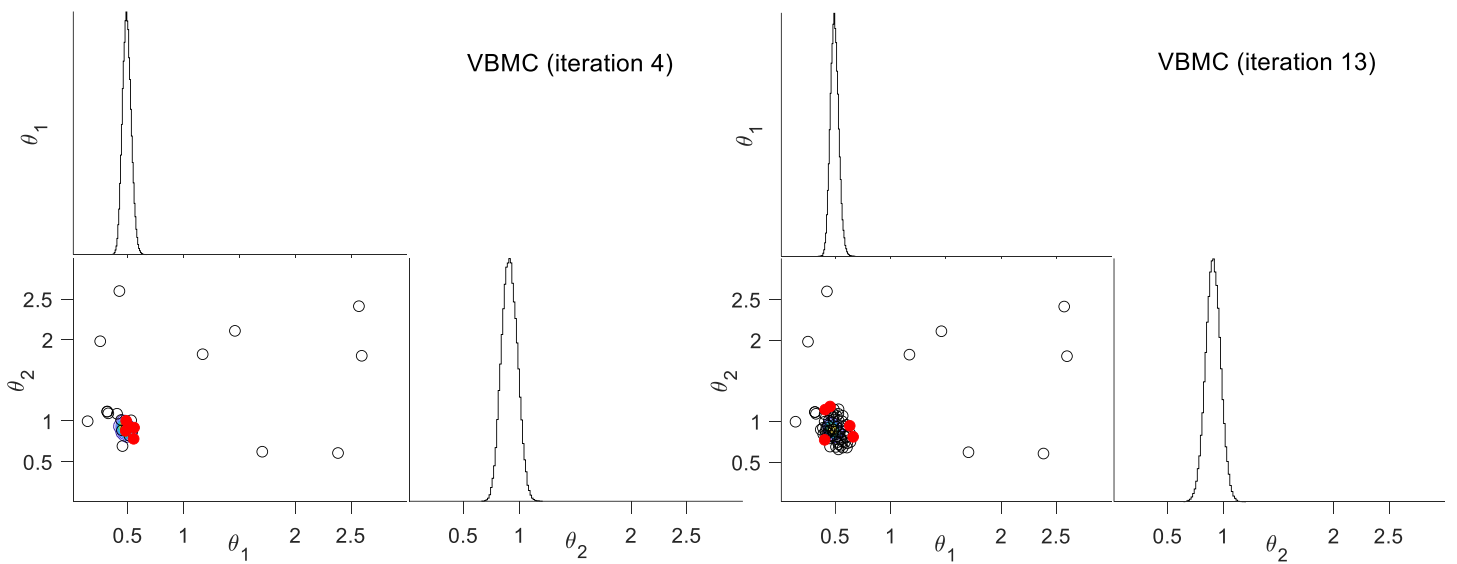

**Fig. 11.** Resulting 1-D and 2-D marginal posterior distributions from VBMC at 4[th] and 13[th] iterations. Red dots indicate samples taken at the current iteration.

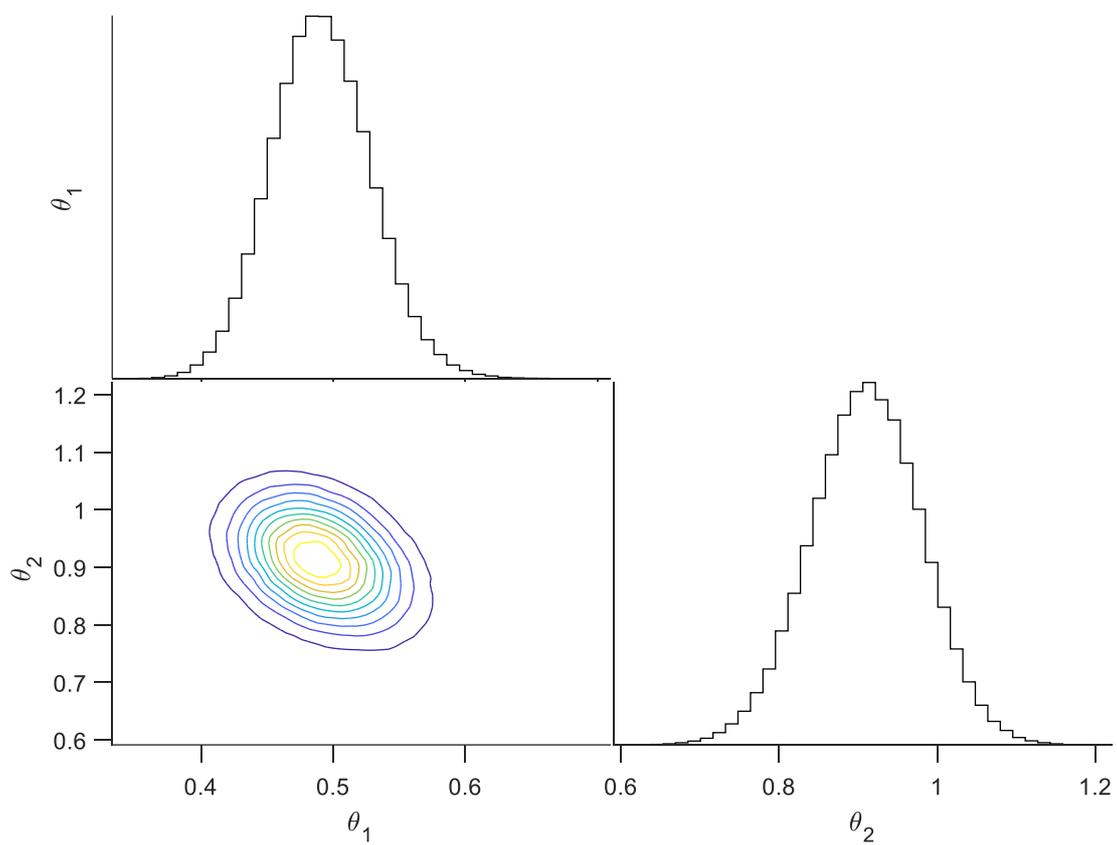

**Fig. 12.** Final 1-D and 2-D marginal posterior distributions from VBMC algorithm.

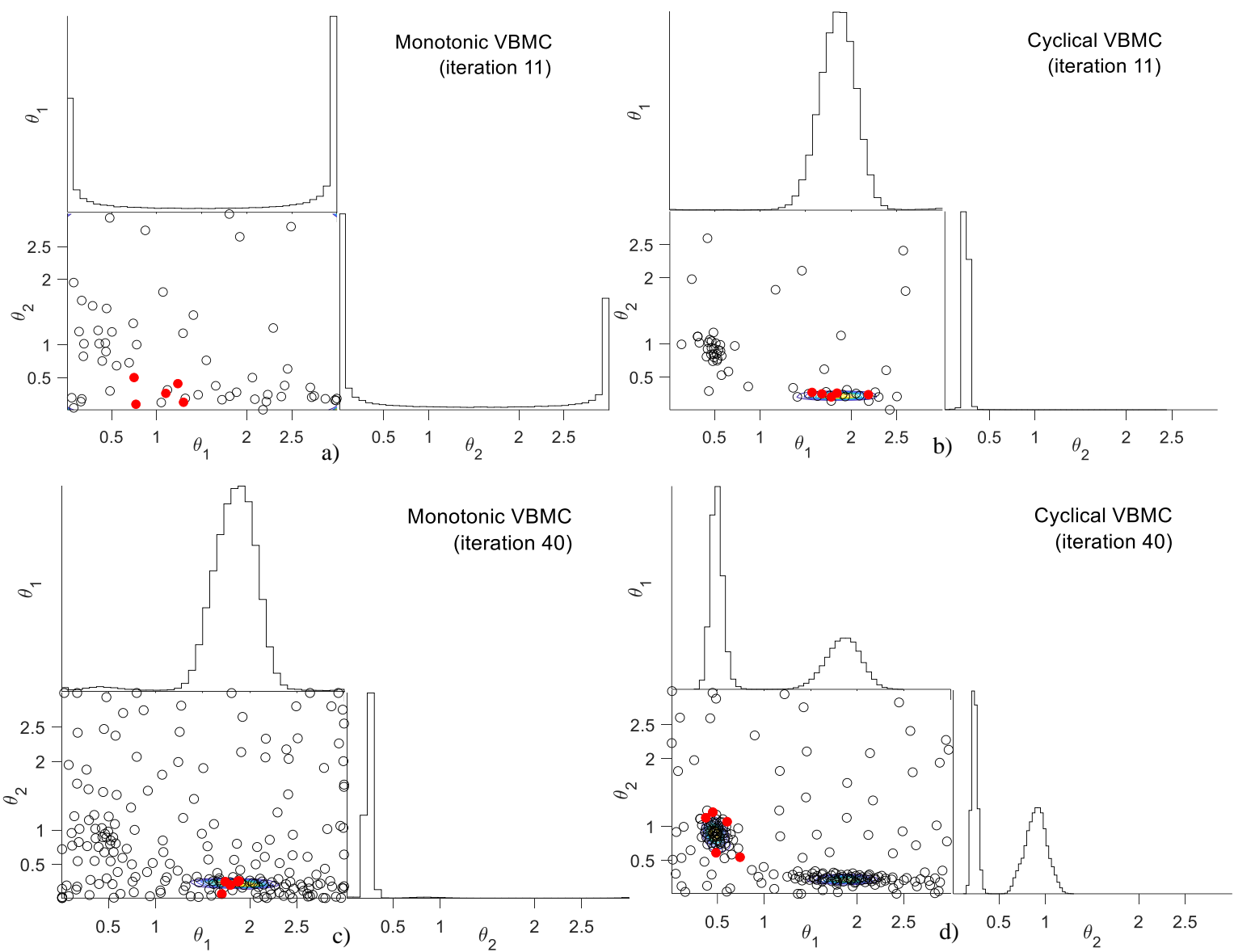

**Fig. 13.** Resulting 1-D and 2-D marginal posterior distributions from VBMC. Red dots indicate samples taken at the current iteration. Black circles indicate samples used for the GP model of the unnormalized posterior at each iteration. a) Monotonic annealing schedule at 11[th] iteration; b) Cyclical Annealing Schedule at 11[th] iteration; c) Monotonic annealing schedule at 40[th] iteration; d) Cyclical Annealing Schedule at 40[th] iteration.

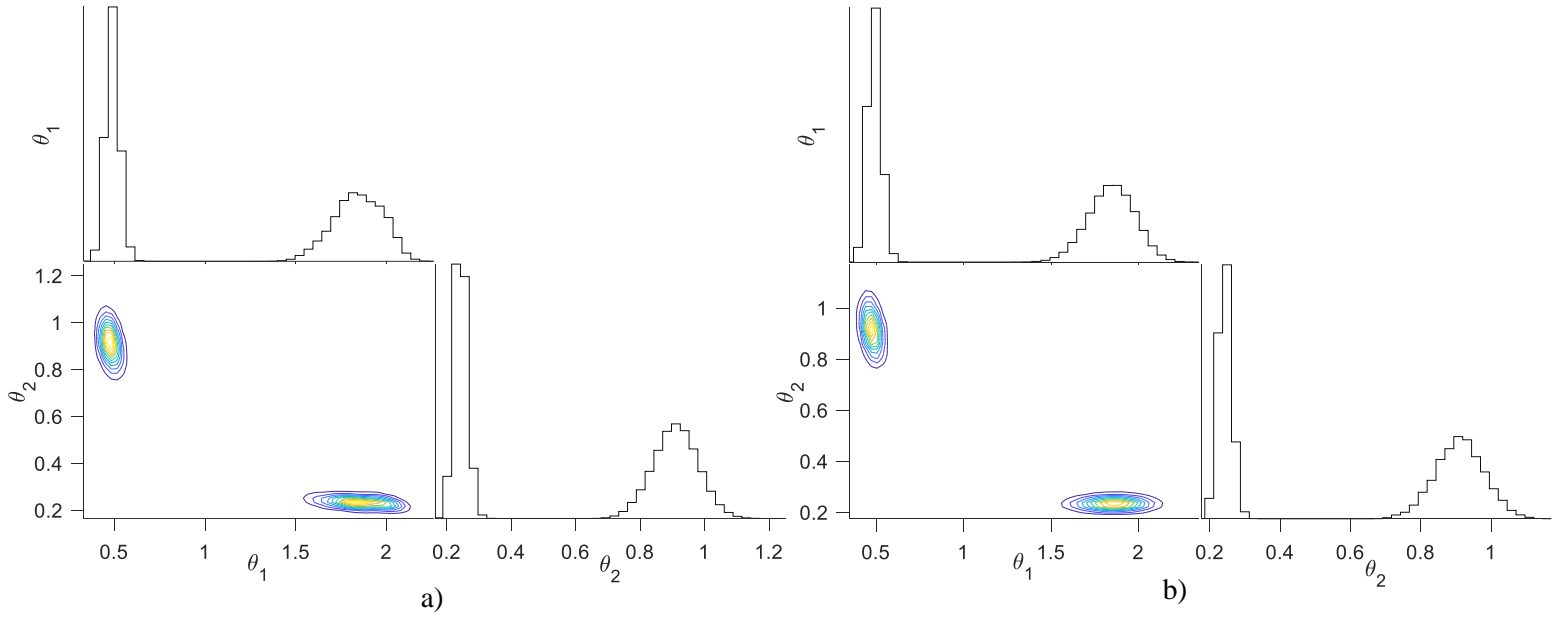

**Fig. 14.** Final 1-D and 2-D marginal posterior distributions from VBMC algorithm. a) Monotonic annealing schedule; b) Cyclical Annealing Schedule.

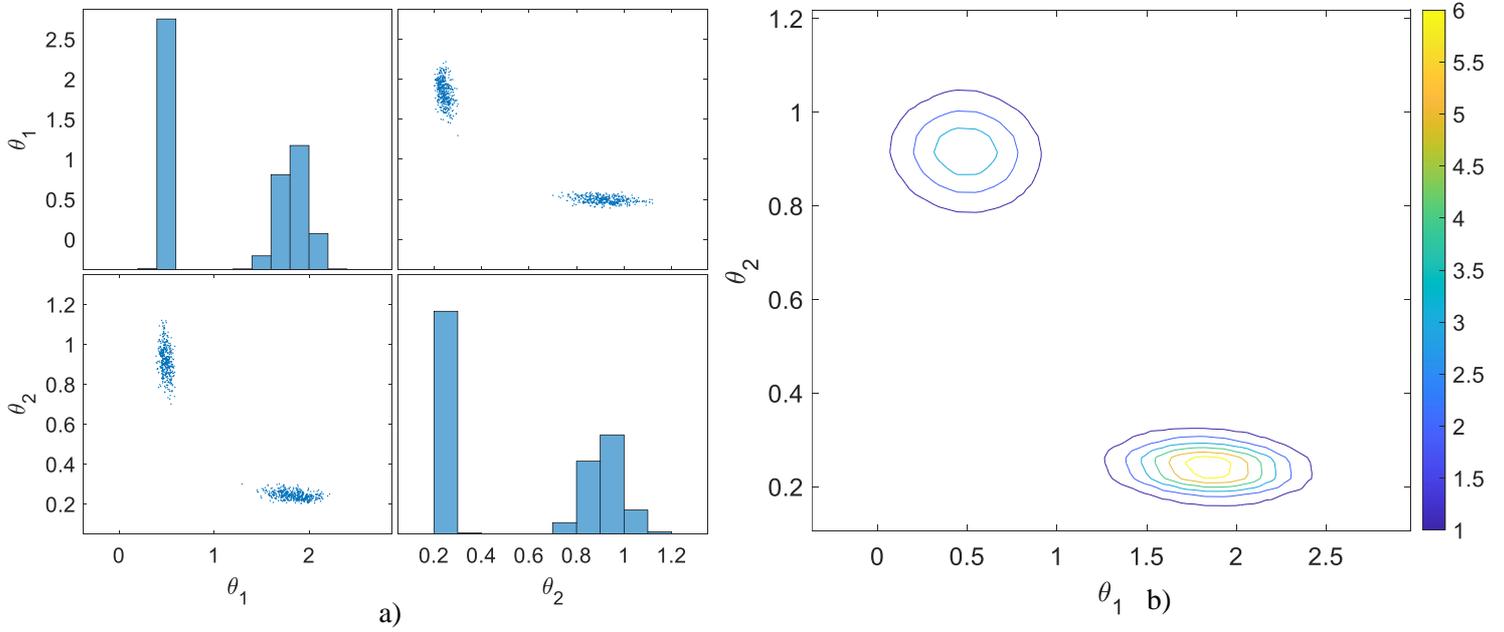

**Fig. 15.** Resulting plots from TEMCMC algorithm. a) Scatterplot; b) 2-D Posterior distribution. The values of the 2-D posterior are given by the numbers on the colour chart.

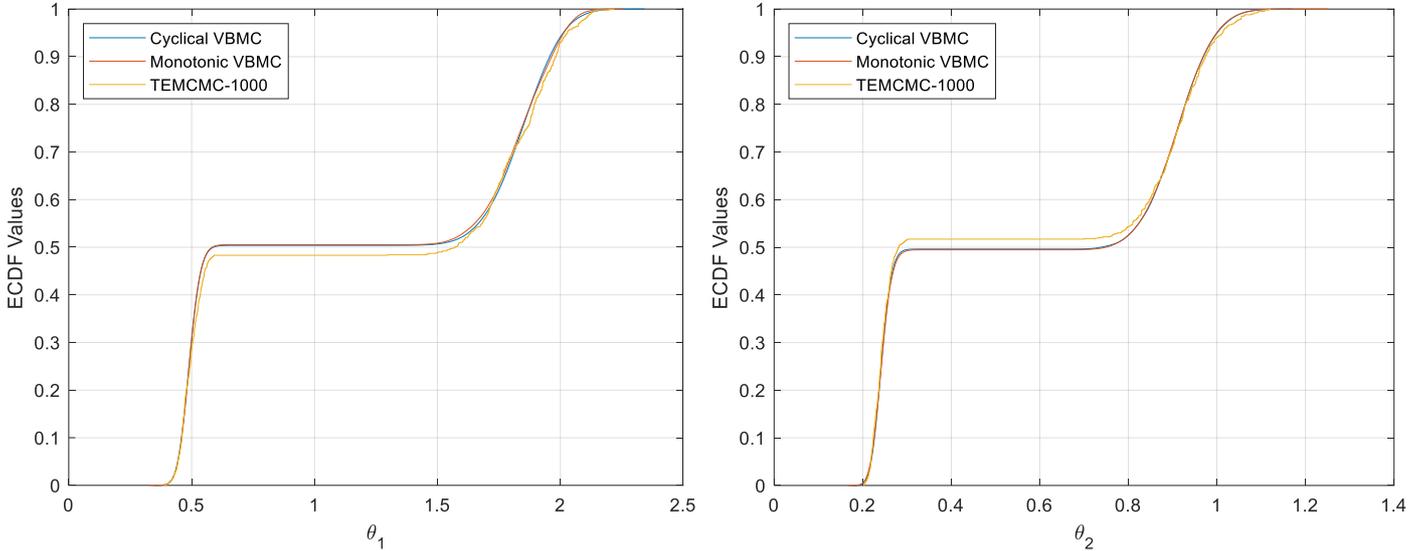

**Fig. 16.** Marginal ECDFs using Cyclical VBMC, Monotonic VBMC and TEMCMC.

**Table 3**

Comparison of numerical results for the mass-spring (multi-modal) system.

| Method | N. of samples | N. of Total Iterations for Convergence | N. of modes found |
| --- | --- | --- | --- |
| VBMC | 70 | 13 | 1 |
| Monotonic VBMC | 275 | 54 | 2 |
| Cyclical VBMC | 260 | 51 | 2 |
| TEMCMC | 5000 | 5 | 2 |

### 4.3 Mass-spring system (unimodal posterior)

In this example taken from [30], for a 4-dimensional Bayesian model updating system, the two Degrees-of-Freedom (DoF) system shown in Fig. 17 is used to compare the performances of all algorithms.

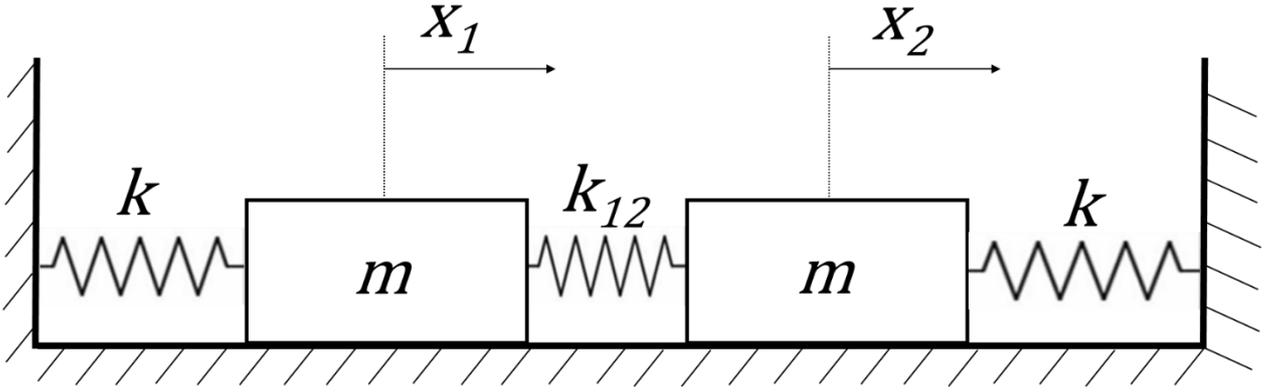

**Fig. 17.** Second mass-spring system.

The 2-DoF system has equal masses $m = 0.5kg$ attached to edge springs with stiffness $k = 0.6N/m$, and the stiffness of the spring between the two masses is $k_{12} = 1N/m$.

The natural frequencies that correspond to the specified properties of the two-degree-of freedom system are given by [30]:

$$\hat{\omega}_1 = \sqrt{\frac{k}{m}} \tag{32}$$

$$\hat{\omega}_2 = \sqrt{\frac{k + 2k_{12}}{m}} \tag{33}$$

The values of the natural frequencies $\hat{\omega}_1$ and $\hat{\omega}_2$ are corrupted with noise as shown below [30]:

$$\omega_1 = \hat{\omega}_1 + \varepsilon_1 \tag{34}$$

$$\omega_2 = \hat{\omega}_2 + \varepsilon_2 \tag{35}$$

Where $\varepsilon_1$ and $\varepsilon_2$ are the noise terms, that follow Gaussian statistical distributions. The mean of both Gaussian distributions is $0Hz$, and their standard deviations are respectively $\sigma_1 = 0.1\hat{\omega}_1 = 0.110Hz$ and $\sigma_2 = 0.1\hat{\omega}_2 = 0.228Hz$. The likelihood function is then given by the following equation [30]:

$$P(y_{obs} | \theta) = \prod_{n=1}^{15} \frac{1}{2\pi\sigma_1\sigma_2} \exp\left[-\frac{(\omega_{1,n} - \hat{\omega}_1)}{2\sigma_1^2} - \frac{(\omega_{2,n} - \hat{\omega}_{2,n})}{2\sigma_2^2}\right] \tag{36}$$

In this example, the parameters $\{k, k_{12}, \sigma_1, \sigma_2\} \equiv \{\theta_1, \theta_2, \theta_3, \theta_4\}$ are assumed to be unknown. The uniform priors $k \sim U(0.1, 4)$ [N/m] and $k_{12} \sim U(0.1, 4)$ [N/m] have been used for the stiffnesses. The prior uniforms taken for the standard deviations $\sigma_1$ and $\sigma_2$ are $\sigma_1 \sim U(10^{-5}, 1)$ [Hz] and $\sigma_2 \sim U(10^{-5}, 1)$ [Hz]. The posterior probability density function of the parameters is updated using the 'experimental measurements', for this example, the fifteen individual experimental 'measurements' of $\omega_1$ and $\omega_2$ used are found in [30].

The final posterior using 1D and 2D marginal distributions are shown in Fig. 18, Fig. 19 and Fig. 20.

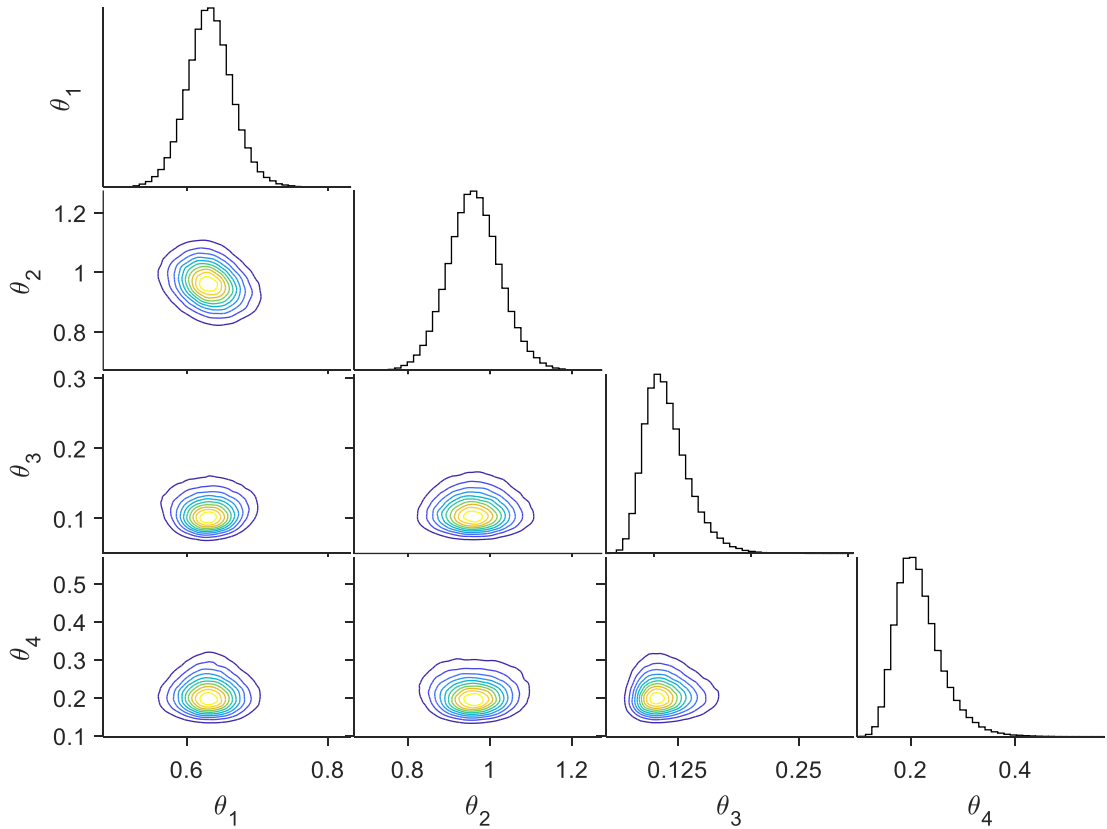

**Fig. 18.** Final 1-D and 2-D marginal posterior distributions from standard VBMC algorithm.

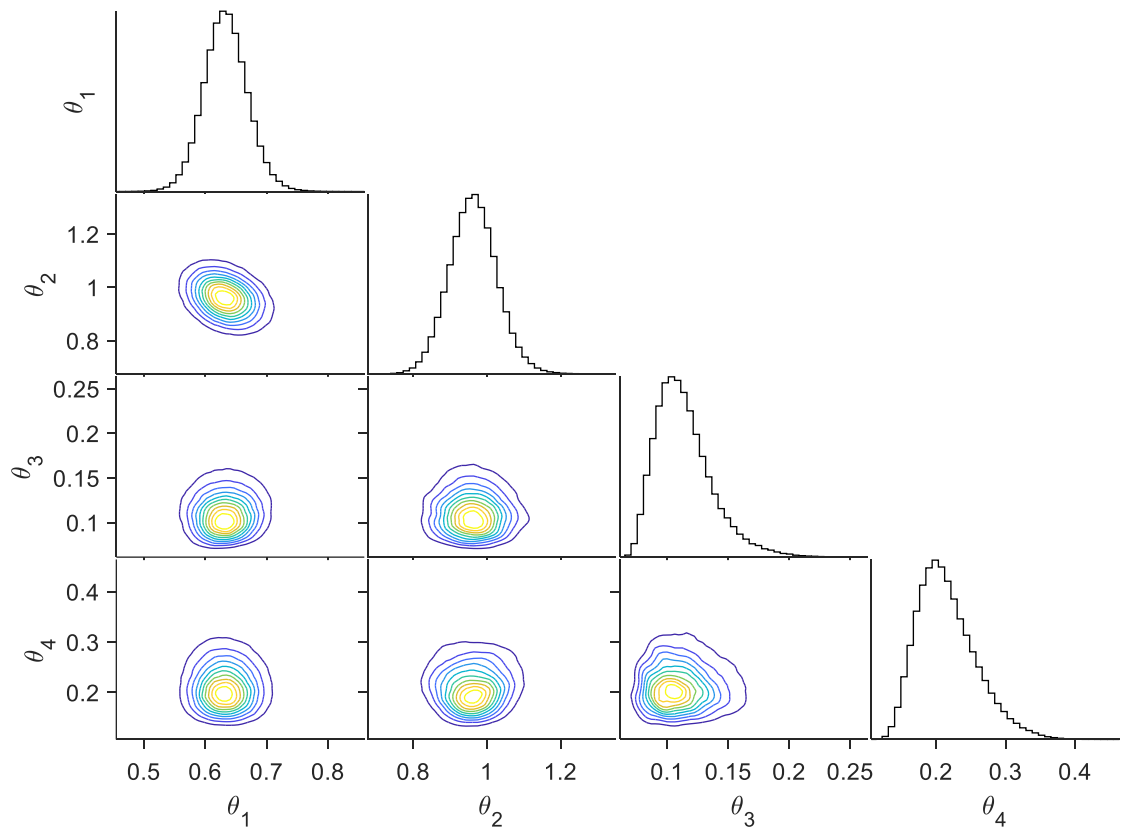

**Fig. 19.** Final 1-D and 2-D marginal posterior distributions from monotonic VBMC algorithm.

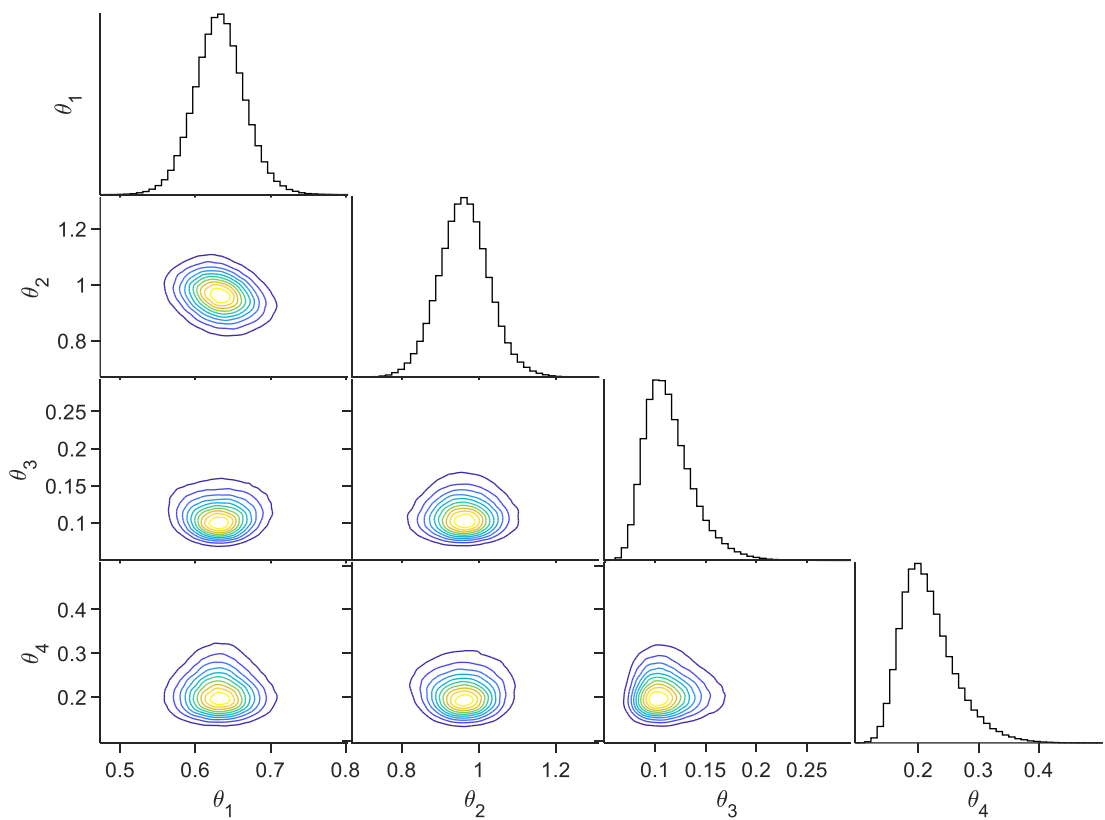

**Fig. 20.** Final 1-D and 2-D marginal posterior distributions from cyclical VBMC algorithm.

**Table 4**

Comparison of numerical results for the mass-spring system (unimodal posterior).

| Method | Sample Mean | Sample C.O.V [%] | N. of samples | N. of Total Iterations for Convergence |
|---|---|---|---|---|
| VBMC | $\begin{bmatrix} 0.633\text{N/m} \\ 0.962\text{N/m} \\ 0.114\text{Hz} \\ 0.217\text{Hz} \end{bmatrix}$ | $\begin{bmatrix} 5.21 \\ 6.67 \\ 20.66 \\ 20.82 \end{bmatrix}$ | 220 | 43 |
| Monotonic VBMC | $\begin{bmatrix} 0.633\text{N/m} \\ 0.963\text{N/m} \\ 0.114\text{Hz} \\ 0.216\text{Hz} \end{bmatrix}$ | $\begin{bmatrix} 5.68 \\ 6.73 \\ 20.70 \\ 19.71 \end{bmatrix}$ | 265 | 52 |
| Cyclical VBMC | $\begin{bmatrix} 0.632\text{N/m} \\ 0.962\text{N/m} \\ 0.114\text{Hz} \\ 0.217\text{Hz} \end{bmatrix}$ | $\begin{bmatrix} 5.34 \\ 6.85 \\ 21.32 \\ 20.85 \end{bmatrix}$ | 260 | 51 |
| TEMCMC [30] | $\begin{bmatrix} 0.625\text{N/m} \\ 1.013\text{N/m} \\ 0.121\text{Hz} \\ 0.229\text{Hz} \end{bmatrix}$ | $\begin{bmatrix} 5.67 \\ 6.80 \\ 17.25 \\ 26.15 \end{bmatrix}$ | 5000 | 5 |
| True Values | $\begin{bmatrix} 0.6\text{N/m} \\ 1\text{N/m} \\ 0.11\text{Hz} \\ 0.228\text{Hz} \end{bmatrix}$ | - | - | - |

Table 4, for a 4D dimensional problem, shows that a significant lower amount of model evaluations is needed for the three VMBC approaches to obtain similar results to the obtained with the TEMCMC algorithm. The results obtained using the TEMCMC algorithm are taken from [30].

## 4.4 Experimental Validation: Aluminium Frame Problem

An aluminium frame experimental case study presented in [30,51], and schematically shown in Fig. 21 is used to compare the four statistical model updating algorithms.

The aluminium frame structure is composed of a total of seven beams. Out of those seven, four beams are vertical (two short internal ones and two long external ones) and three beams are horizontal (all with identical length). Two masses $m_1$ and $m_2$ of variable position also form part of the structure. Each of those two masses is attached to one of the two short vertical beams at distances from the bottom of the beam $pm_1$ and $pm_2$. The movable masses are used in the structure to induce an effect in the modal properties of the system comparable to the one induced by possible structural damages.

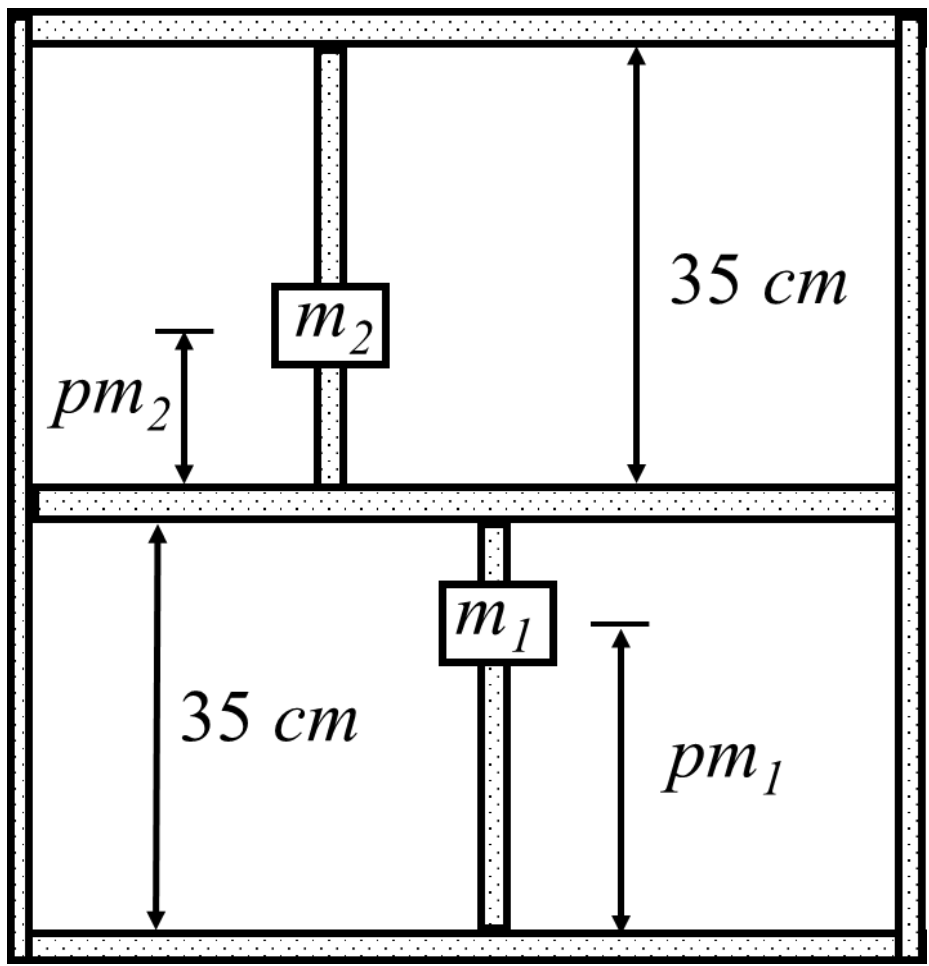

**Fig. 21.** Schematic representation of aluminium frame with moveable masses (adapted from [30]).

Experimental data was used to obtain the first six natural frequencies of the structure for eleven different combinations $\{pm_1, pm_2\}$. A summary of the results is found on Table 5.

**Table 5**

Experimental results (data from [52])

| Mode | | 1 | 2 | 3 | 4 | 5 | 6 |
|---|---|---|---|---|---|---|---|
| Modal shape | | 1st in-plane bending mode | 1st out-of-plane bending mode | 1st torsional mode | 2nd in-plane bending mode | 2nd out-of-plane bending mode | 2nd torsional mode |
| Exp. | Positions (pm1,pm2) [cm] | \multicolumn{6}{c}{Frequencies [Hz]} | | | | | |
| | | $\omega_1$ | $\omega_2$ | $\omega_3$ | $\omega_4$ | $\omega_5$ | $\omega_6$ |
| 1 | (5, 5) | 20.11 | 22.79 | 47.52 | 63.96 | 183.82 | 283.51 |
| 2 | (5, 20) | 18.72 | 20.46 | 46.97 | 72.24 | 214.84 | 296.32 |
| 3 | (5, 35) | 17.72 | 18.29 | 46.42 | 63.45 | 196.38 | 278.70 |
| 4 | (20, 5) | 19.40 | 22.39 | 46.32 | 61.78 | 173.49 | 259.76 |
| 5 | (20, 20) | 17.91 | 20.28 | 45.67 | 64.73 | 190.84 | 284.09 |
| 6 | (20, 35) | 16.71 | 18.21 | 45.18 | 56.53 | 177.97 | 264.44 |
| 7 | (35, 5) | 17.71 | 21.76 | 44.00 | 59.48 | 164.05 | 254.48 |
| 8 | (35, 20) | 16.91 | 19.82 | 43.15 | 60.06 | 175.75 | 279.10 |
| 9 | (35, 35) | 15.95 | 17.89 | 42.44 | 50.66 | 163.55 | 257.82 |
| 10 | (11, 11) | 19.58 | 21.73 | 47.00 | 67.54 | 196.21 | 285.95 |
| 11 | (29, 29) | 16.65 | 18.85 | 43.93 | 55.43 | 174.35 | 284.84 |

Following the same approach used on [30], a surrogate model of the expensive-to-evaluate Finite Element Model (FEM) of the aluminium frame structure is built. An Artificial Neural Network (ANN) is trained using a set of simulated values of the expensive-to-evaluate FEM from the database [52]. The architecture for the ANN and calibrated model of [30] is also used. The ANN's architecture includes three layers: input, hidden and output with respectively two, ten and six nodes. Further details on the calibration procedure employed can be found in the paper [30].

The set of input data presented in the form of a scatterplot are shown in Fig. 22.

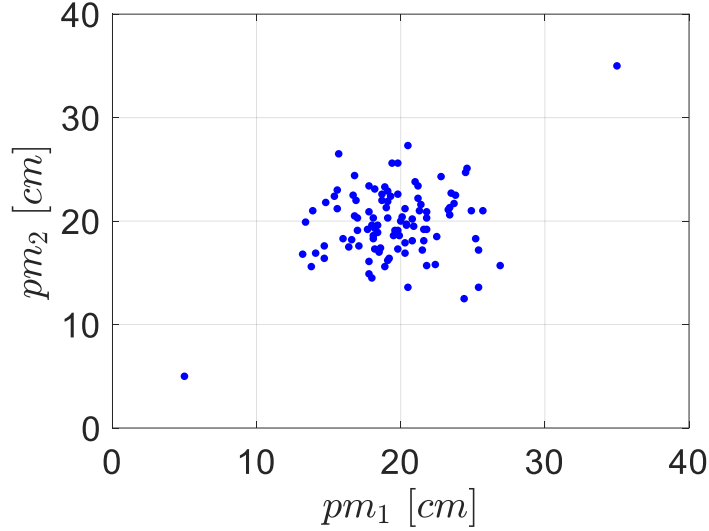

**Fig. 22.** Set of input data used for calibration of the ANN.

In this example, the parameters $\{pm_1, pm_2, \sigma_1, \ldots, \sigma_6\} \equiv \{\theta_1, \theta_2, \theta_3, \ldots, \theta_8\}$ are assumed to be uncertain and independent from each other. The uniform priors $pm_1 \sim U(5,35)$ [cm] and $pm_2 \sim U(5,35)$ [cm] have been used for the position of the masses. The uniform prior taken for all the measurement 'noises' $\sigma_v$ corresponding to the natural frequency $\omega_v$ (for $v=1,\ldots,6$) are $\sigma_v \sim U(10^{-3}, 100)$ [Hz].

The overall likelihood function $LF$ is given by the weighted addition of the three likelihood functions $lf_1, lf_2, lf_3$ defined below [30,51]:

$$lf_1 = \prod_{v=1}^{6} \frac{1}{\sigma_v \sqrt{2\pi}} \exp\left[-\frac{(\omega_v - M_v)^2}{2\sigma_v^2}\right] \tag{37}$$

$$lf_2 = \prod_{v=1}^{6} 1 - \exp\left[-\frac{1}{(\omega_v - M_v)^2}\right] \tag{38}$$

$$lf_3 = \prod_{v=1}^{6} 1 - \exp\left[-\sqrt{\frac{1}{(\omega_v - M_v)^2}}\right] \tag{39}$$

$$LF = \frac{1}{3}\sum_{p=1}^{3} lf_p \tag{40}$$

Where $M_v$ is the model output for $\omega_v$.

In this example, only the last two experiments (10 and 11) of Table 5 were used as observations in the Bayesian model updating framework. These two experiments were selected to recreate an example where during an observation period, two different states of damage might occur. Therefore, these two different states of damage of the structure will have distinct natural frequencies that correspond to different values of the uncertain physical parameters to be inferred.

Fig. 23, Fig. 24, Fig. 25 and Fig. 26 show the final 1-D and 2-D marginal posterior distributions obtained from the algorithms. The modes of the posterior are not found at the positions of the masses on the experiments 10 and 11. This is due to two reasons: using an ANN that had not been trained with enough samples covering the entire parameter space as shown in Fig 22, and the fact that the experiments do not bring enough information to make the Bayesian model updating problem globally identifiable. As a result, a highly multi-modal posterior of great interest to test the applicability of the three VBMC algorithms is found.

From Fig. 23, it can be observed that the standard VBMC is unable to find one of the modes shown in the marginal posterior of $\theta_1$. However, this missing mode on Fig. 23 is clearly seen in the monotonic VBMC and cyclical VBMC of Fig. 24 and Fig. 25 respectively. It can also be seen in Fig. 25 that the cyclical VBMC posterior has the highest degree of similarity to the posterior obtained in Fig. 26 by the TEMCMC algorithm. It is also found that the three VBMC approaches face slight difficulties to obtain an approximation to the posterior compared to the previous numerical examples. This is due to the very multi-modal posterior found in this problem, where the greatest difficulty is found in the approximation of the tail behaviour of the true posterior.

In Fig. 27 the ECDFs for the cyclical VBMC, monotonic VBMC, VBMC and TEMCMC are shown. The resulting ECDFs are found to be similar to the one obtained by the TEMCMC, with the cyclical VBMC showing the highest similarity.

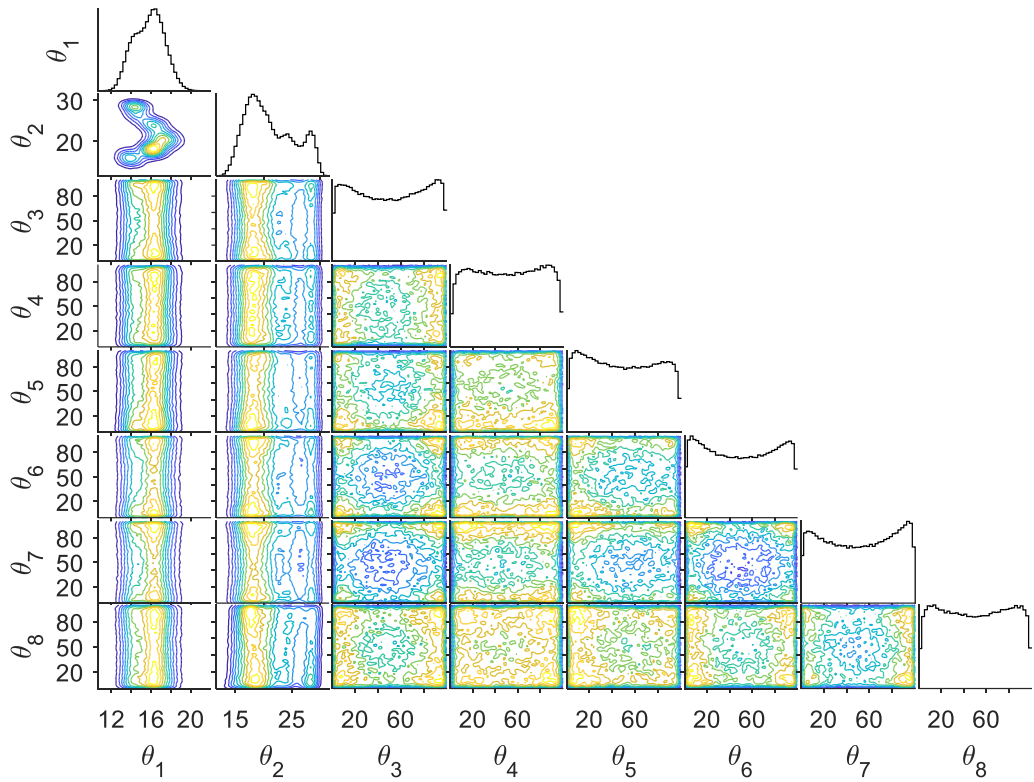

**Fig. 23.** Final 1-D and 2-D marginal posterior distributions from VBMC algorithm.

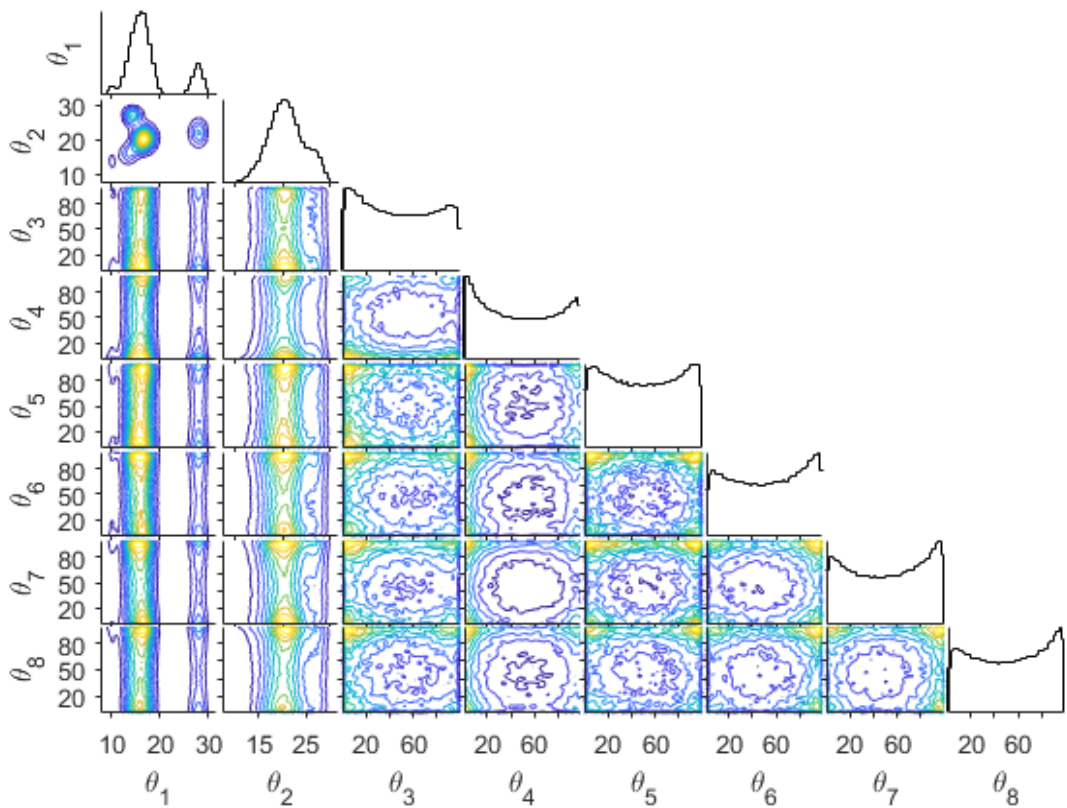

**Fig. 24.** Final 1-D and 2-D marginal posterior distributions from monotonic VBMC algorithm.

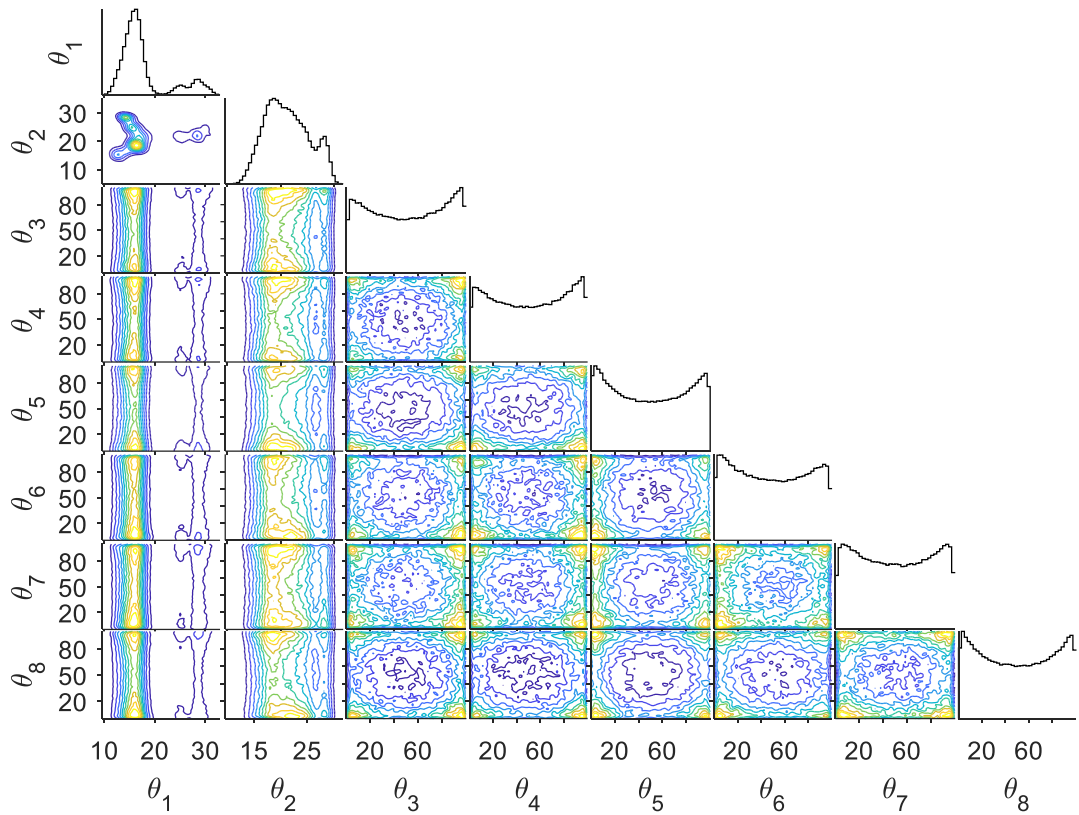

**Fig. 25.** Final 1-D and 2-D marginal posterior distributions from cyclical VBMC algorithm.

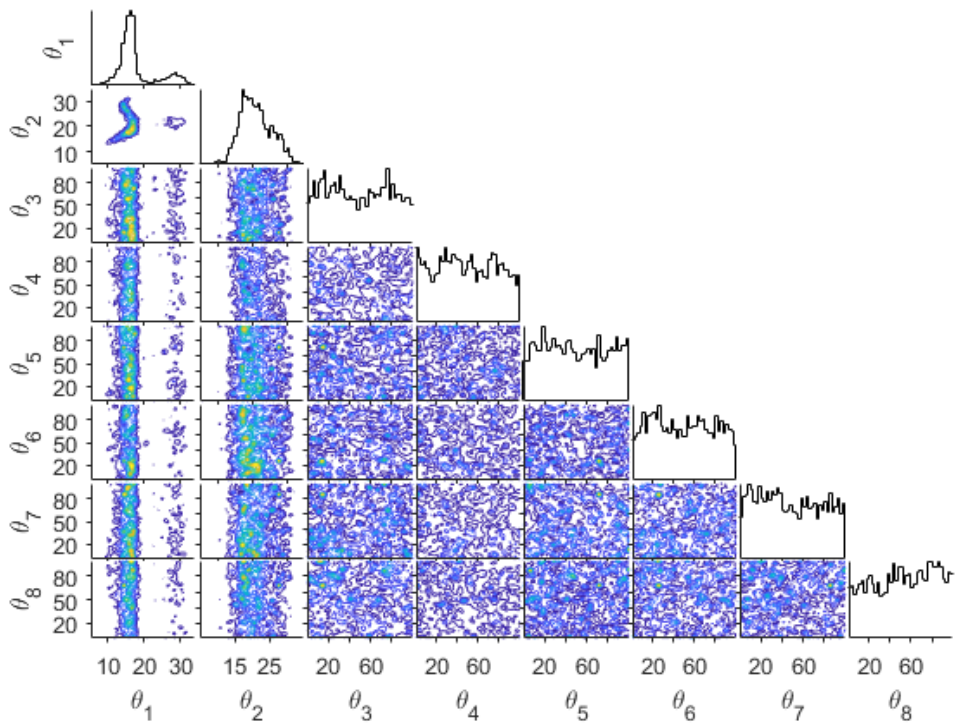

**Fig. 26.** Final 1-D and 2-D marginal posterior distributions from TEMCMC.

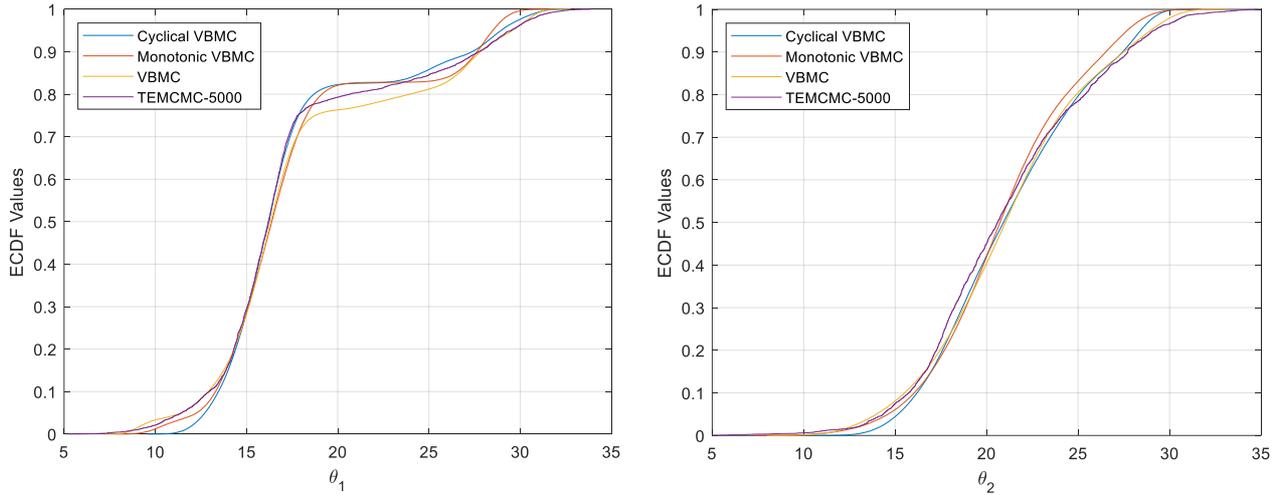

**Fig. 27.** Marginal ECDFs using Cyclical VBMC, Monotonic VBMC, VBMC and TEMCMC (5000 samples).

Table 6 compares the results obtained using the three VBMC approaches and TEMCMC. It can be seen that for a significantly lower amount of model evaluations, all the three VBMC algorithms show similar results to the TEMCMC algorithm.

**Table 6**

Comparison of numerical results for the multi-modal aluminium frame problem.

| Method | N. of samples | N. of Total Iterations for Convergence |
|---|---|---|
| VBMC | 135 | 26 |
| Monotonic VBMC | 270 | 53 |
| Cyclical VBMC | 265 | 52 |
| TEMCMC | 20000 | 4 |

**4.5 Discussion of Results**

The results obtained for the case studies illustrated in section 4 indicate that:

The standard VBMC algorithm displays an excellent performance for the unimodal posterior distribution example, however, for the three multi-modal posterior examples it gets stuck at the initially found mode. This is due to the nature of the algorithm: the active sampling used is unable to escape from that mode. In other words, the algorithm proceeds to only sample in the vicinity of that found mode due to its exploitation nature. A better approximation of the posterior may have been found if a higher number of function evaluations had been used for the initial training set, as that would have meant a better exploration of regions with high probability density. The disadvantage of running a higher number of function evaluations would be that as no guided exploration is used, a significant number of those function evaluations performed would be wasted.

For the three examples featuring the multi-modal posterior distributions, it was shown that the cyclical VBMC outperforms the other methods: it gradually improves convergence by reopening paths and by leveraging on the previous cycles as warm re-starts. When the monotonic VBMC algorithm is used, the paths are formed throughout the iterations, but they are not reopened, meaning that once the exploration phase has finished, exploitation of the previously found regions of high probability density occurs.

The second numerical example focuses on a mildly multi-modal distribution where the difference between the monotonic and cyclical VBMC results is expected to be modest.

In the other multi-modal examples, it was found that to obtain ECDFs of similar accuracy to the TEMCMC algorithm, the cyclical VBMC algorithm required a significantly lower amount of function evaluations of the model (at least an order of magnitude lower). It can be concluded that the annealing schedule improves the convergence of the approximated posterior to the true posterior. This improvement is at the expense of a slight additional computational cost compared to the standard VBMC when dealing with unimodal posterior distributions. This extra computational cost is certainly justified when the interest is the accurate evaluation of highly multi-modal posteriors which are expected to be found in the Bayesian model updating problems of engineering applications. This is certainly of great interest for methods that require the evaluation of cumulative density functions, such as reliability analysis techniques.

# 5 Conclusions

In this paper, an approach based on variational inference for the estimation of the multi-modal posterior distributions of the latent parameters of an expensive-to-evaluate physics-based model, given available data, has been proposed. The proposed cyclical VBMC approach yields a non-parametric estimation of the posterior distribution of the identified parameters by combining the active-sampling Bayesian quadrature with a Gaussian-process based variational inference. Multi-modal smooth posteriors can be captured as it uses a multivariate Gaussian mixture postulated posterior. Variational whitening is also used in this proposed approach for a more accurate posterior approximation. The cyclical VBMC algorithm overcomes the constraints raised by poor initializations when the number of model runs that can be explored is small. This is done employing an artificial temperature parameter that cyclically anneals the unnormalized posterior, improving the mode coverage and exploration abilities of the procedure. Three numerical examples and one experimental investigation have shown the advantages of the cyclical VBMC when dealing with multi-modal posteriors and a limited number of physics-based model runs.

The proposed cyclical VBMC approach may benefit other engineering applications, including Bayesian Experimental Design [53–55], and optimal sensor placement frameworks based on information theory [56,57], since it may reduce the computational cost required for the statistical model updating part of these approaches when dealing with less than 20 uncertain parameters. These applications are currently being explored, with particular interest to the cases with more than 20 uncertain parameters, which are often encountered in engineering.

**Acknowledgements**

Felipe Igea and Alice Cicirello thanks the EPSRC and Schlumberger for an Industrial Case postgraduate scholarship (Grant ref. EP/T517653/1).